\documentclass[twocolumn,showpacs,preprintnumbers,amsmath,amssymb]{revtex4}
\usepackage{amsmath,amstext,amsthm,vmargin,enumerate,graphicx}
\usepackage{subfigure}

\newcommand{\ra}{\langle r\rangle}

\begin{document}

\title{Self-organized criticality in the intermediate phase of rigidity
percolation}

\author{M.-A. Bri\`ere}
\email{marc-andre.briere@umontreal.ca}
\affiliation{D\'epartement de physique and Regroupement qu\'eb\'ecois sur les
mat\'eriaux de pointe, Universit\'e de Montr\'eal, case postale 6128,
succursale centre-ville, Montr\'eal (Qu\'ebec), Canada H3C 3J7}
\author{M.V. Chubynsky}
\email{mykyta.chubynsky@umontreal.ca}
\affiliation{D\'epartement de physique and Regroupement qu\'eb\'ecois sur les
mat\'eriaux de pointe, Universit\'e de Montr\'eal, case postale 6128,
succursale centre-ville, Montr\'eal (Qu\'ebec), Canada H3C 3J7}
\author{Normand Mousseau}
\email{normand.mousseau@umontreal.ca}
\affiliation{D\'epartement de physique and Regroupement qu\'eb\'ecois sur les
mat\'eriaux de pointe, Universit\'e de Montr\'eal, case postale 6128,
succursale centre-ville, Montr\'eal (Qu\'ebec), Canada H3C 3J7}

\date{\today}

\begin{abstract}

Experimental results for covalent glasses have highlighted the existence of a
new self-organized phase 
due to the tendency of glass networks to minimize internal stress. Recently, we
have shown that an equilibrated self-organized two-dimensional lattice-based
model also possesses an intermediate phase in which a percolating rigid
cluster exists with a probability between zero and one, depending on
the average coordination of the network. In this paper, we study the
properties of this intermediate phase in more detail. We find that microscopic
perturbations, such as the addition or removal of a single bond, can affect
the rigidity of macroscopic regions of the network, in particular, creating or
destroying percolation. This, together with a power-law distribution of rigid
cluster sizes, suggests that the system is maintained in a critical state on the
rigid/floppy boundary throughout the intermediate phase, a behavior similar to
self-organized criticality, but, remarkably, in a thermodynamically
equilibrated state. The distinction between percolating and
non-percolating networks appears physically meaningless, even though the
percolating cluster, when it exists, takes up a finite fraction of the network.
We point out both similarities and differences between the
intermediate phase and the critical point of ordinary percolation models
without self-organization. Our results are consistent with an interpretation
of recent experiments on the pressure dependence of Raman frequencies in
chalcogenide glasses in terms of network homogeneity.

\end{abstract}


\maketitle

\section{Introduction}

The concept of rigidity percolation, first introduced about 25 years ago by
Thorpe~\cite{thorpe83} based on work by Phillips~\cite{phillips79}, describes
how an elastic network goes from floppy to rigid as constraints are added to
it. This theory has been applied with success to many systems, including
covalent glasses~\cite{thorpe00, travbool, travthorpe} and
proteins~\cite{proteins01}. In the last decade, however, experimental studies
have shown that the rigidity phase diagram could be more complex than
initially thought, uncovering the presence of an intermediate phase between
the floppy phase and the stressed-rigid phase, with the system in the
intermediate phase being rigid but unstressed~\cite{travbool, bool00sise,
bool01rev, bool01gese, bool02rev, bool02cr, bool03pse, bool03geasse, bool04geps,
bool05pgese, bool05geass, bool05geass2, bool05press, bool05sil,lucov4}.

A basic explanation for this new phase was first proposed by Thorpe {\it et
al.}~\cite{thorpe00}. It was shown that when the network self-organizes in
order to minimize the stress, the rigid but unstressed intermediate phase can
indeed arise.

In the original work by Thorpe and collaborators~\cite{thorpe00, czech01}, as
well as subsequent simplified models of chalcogenide glasses by Micoulaut and
Phillips~\cite{micoulaut02, micoulaut03}, networks were constructed using an
``aggregation'' process, in which bonds or simple network units were added to
the network without subsequent equilibration. More recently, Barr\'e {\it et
al.}~\cite{barre} have considered a thermodynamically proper model with an energy cost
associated with stress and showed that in the canonical ensemble, the
intermediate phase still exists. In a recent paper, we have confirmed this
result for the $T=0$ version of the model of Barr\'e {\it et al.}, but on a more
realistic regular lattice, and also have shown that the intermediate phase is
entropically feasible in actual physical systems. In both the model by Barr\'e
{\it et al.} and our model, the intermediate phase has an unusual property:
both percolating and non-percolating networks coexist in the ensemble at all
mean coordination numbers within the intermediate phase.

While providing a general support for self-organization, these previous
studies did not look in detail at the properties of networks in the
intermediate phase. Here we provide a first glimpse at some of these
properties. First, we show that in both percolating and non-percolating
networks, the sizes of non-percolating clusters have a power-law distribution.
In effect, the system remains in a critical state over an extended range of
mean coordinations, corresponding to a self-organized
critical phase~\cite{bak2}, but in an equilibrium system. Second, we find that
adding or removing a single bond affects the rigidity of macroscopic parts of
the network in the intermediate phase and in particular, can turn a
non-percolating network into a percolating one and vice versa. This property supports the interpretation by Wang and co-workers~\cite{bool05press}
of the puzzling response of
vibrational frequencies to applied pressure that was observed in their
experiments. Using our results, we address some intriguing
questions that have to do with the unusual coexistence of percolating and
non-percolating networks in the intermediate phase. In particular, we show that
percolating and non-percolating networks can be considered identical in the
intermediate phase, as both stay on the edge of percolation. 

This paper is structured as follows. In the next section, we briefly review
the intermediate phase. We then present our methodology. In the fourth
section, we present our results on the properties of rigid clusters (both
percolating and non-percolating) in our model. In Section V, we look at the
response of the network to local perturbations. In Section VI, we discuss
how our results can help understand the experiments by
Wang {\it et al.}~\cite{bool05press}. We finish with conclusions. 

\section{The intermediate phase}

Using the mean-field approach first introduced by Maxwell~\cite{maxwell} and known
as Maxwell counting, we can define the rigidity of a network in terms of its
number of zero-frequency motions, or {\it floppy modes}, $F$. In a
$d$-dimensional network, each atom has $d$
degrees of freedom. In an unconstrained network of $N$ atoms, each degree of
freedom corresponds to a floppy mode and thus $F=dN$. Assuming that each
added constraint takes away one floppy mode, we can write
\begin{equation}
F = dN - N_c,\label{eq:1}
\end{equation}
where $N_c$ is the number of constraints in the network. As $N_c$ increases,
$F$ decreases, until $F=0$ is reached, and then there are no floppy modes left
and the network undergoes a rigidity transition from floppy to rigid
where a percolating rigid cluster emerges in the network (rigid clusters
are sets of mutually rigid atoms)  Disordered networks can be conveniently
characterized by the {\it mean coordination} $\ra$, which is the average
number of bonds connecting a site. In the mean-field approximation, the location
of the rigidity transition only depends on $\ra$ and not on other details, such
as fractions of sites with a particular coordination. The transition occurs at
$\ra = 4$ for the triangular lattice of elastic springs and $\ra=2.4$ in
chalcogenide glasses; in the latter case, it is assumed that both
bond-stretching and bond-bending constraints are taken into account.

To go beyond the mean-field theory, corrections must be made. For example,
adding a constraint to an already rigid region does not remove a floppy mode.
Such type of constraint is called redundant. Redundant constraints introduce
{\it stress} into the network. Such constraints do not change
the number of floppy modes (and thus violate the assumption of the Maxwell
counting); taking this into account, Eq.~(\ref{eq:1}) becomes
\begin{equation}
F = dN - (N_c - N_r),
\end{equation}
where $N_r$ is the number of redundant constraints. Another type of correction
is due to the fact that even above the rigidity transition there can still be
some floppy inclusions and thus $F>0$ at the transition.

To find $N_r$, Jacobs and co-workers have introduced a topological
algorithm, the \emph{pebble game}~\cite{jacobs95,jacobs97}. This algorithm is
based on a theorem by Laman~\cite{laman} which states that in two dimensions a
generic network with $N$ sites and $B$ bonds does not have
a redundant bond if and only if no subset of the network containing $n$ sites
and $b$ bonds violates $b\le 2n-3$. A similar approach
works in 3D, but in general only for bond-bending networks like those used to
model chalcogenide glasses.

The pebble game, described in more detail in Ref.~\cite{jacobs97},
characterizes the global rigidity of a network, provides its complete
decomposition into rigid clusters and finds stressed regions.
The approach uses only the topology of the network and not its exact geometry.
Using the pebble game, it was possible to show that the rigidity transition
occurs at $\ra =3.961$ $\pm$0.001~\cite{jacobs95} for the central-force
triangular lattice and at $\ra\approx 2.385$~\cite{thorpe00} for
an amorphous bond-bending 3D network.

While early measurements also identified a rigidity transition in these
glasses near $\ra=2.4$, recent experiments have shown that there is not one
but two transitions~\cite{travbool, bool00sise,
bool01rev, bool01gese, bool02rev, bool02cr, bool03pse, bool03geasse, bool04geps,
bool05pgese, bool05geass, bool05geass2, bool05press, bool05sil}.
Starting at low mean coordination, the first transition seems to go from a
floppy to a rigid but unstressed phase, and the second one on to a rigid and
stressed phase. This new rigid but unstressed phase is known as the
\emph{intermediate phase} and is believed to be related to the
self-organization of the system in order to minimize the stress in the
network.

This interpretation is supported by a number of theoretical works. In their
pioneering work, Thorpe and collaborators demonstrated that a network constructed by
adding bonds with no stress allowed until it becomes inevitable would go through
three phases: a floppy, a rigid-unstressed and a rigid-stressed
phases~\cite{thorpe00}. The first transition, between the floppy and the
rigid-unstressed phases, is the {\it rigidity transition}; the second one, between
the rigid-unstressed and rigid-stressed phases, is the {\it stress transition} and
happens immediately when avoiding stress is no longer possible, The intermediate
phase survives when the rigid but stress-free networks are fully equilibrated as was
demonstrated by Barr\'e and co-workers on Bethe lattices~\cite{barre} and Chubynsky
\emph{et al.}~\cite{chubynsky06} on two-dimensional triangular lattices. This is not
the only approach to self-organization, however, and Micoulaut and
collaborators~\cite{micoulaut02, micoulaut03} have shown that it is possible to
recover an intermediate phase in a stressed network if this stress is localized.

\section{Methodology}

The model we study here is the same as the one used in our previous
paper~\cite{chubynsky06}. In our simulations, we use the pebble game algorithm
described in the previous section; our computer code is based on the original
program by D.J.~Jacobs and M.F.~Thorpe.

We consider 2D triangular bond-diluted central-force networks. While we cannot
make a direct comparison
with experiment, previous work has shown that the triangular lattice presents
an intermediate phase similar to that of covalent glasses with angular
constraints; our results should therefore be applicable, at least
qualitatively, to experiments.

In the original model of Thorpe and collaborators~\cite{thorpe00}, bonds were
added one by one and each checked for redundancy; redundant (stress-causing) bonds were rejected. Each new added bond was frozen in the
network and was never moved nor removed. This procedure does not guarantee that
stress-free networks are equiprobable, making some networks more likely than
others. To eliminate this bias, we introduced a bond-equilibration scheme,
allowing the system to rearrange itself by moving bonds around. Each time a
new bond is added, bonds are reshuffled throughout the lattice: a bond is
picked at random, removed, and then a new bond is inserted in a random place
choosing among those where it would not create stress. This bond-shuffling
procedure is repeated until the system is equilibrated. We find that an
equilibration of 10 iterations per added bond below $\ra$=3.5 and 100
iterations above $\ra$=3.5 is enough for convergence (for more details, see
Chubynsky {\it et al.}~\cite{chubynsky06}) and it is the equilibration scheme
we use throughout this paper, unless stated otherwise. In equilibrium, all stress-free networks with a given
number of bonds (or mean coordination) are equiprobable. This corresponds to the
thermodynamic equilibrium at $T\to 0$ for any model in which all stress-free
networks have equal energy, but the energies of stressed networks are higher.

As discussed in our previous paper~\cite{chubynsky06}, the intermediate phase
in the model described above is associated with a non-trivial probability of
finding a percolating network in this equilibrated model: this probability
rises linearly from zero at the rigidity transition ($\ra\approx 3.945$) to
one at the stress transition ($\ra=4.0$), a result similar to that obtained
by Barr\'e {\it et al.} on the Bethe lattice~\cite{barre}. Thus there
are both percolating and non-percolating networks and these two classes need
to be studied separately.

All results presented in this paper (with the exception of the cluster size
distributions given in Fig.~\ref{fig:clst_dist}) are obtained by running 200
independent simulations and obtaining the quantities of interest at different
$\ra$. When overall averages are presented, these are obtained by averaging
over all these simulations. When, e.g., an average for percolating networks is
presented, then the averaging is done over only those of these networks that are
percolating at the given $\ra$. Obviously, close to the rigidity transition,
very few of the 200 networks are percolating, and so the corresponding quantity
will be an average over a very small number of realizations and may contain a
bigger error. When we report the results of insertion or removal of a single
bond, only one attempt of insertion/removal per network is made, unless
stated otherwise.

\section{Properties of rigid clusters}

In this section, we look at properties of rigid clusters (both percolating and
non-percolating) that exist in self-organized networks in the intermediate
phase. Some properties of rigid clusters that distinguish them from, e.g.,
clusters in usual connectivity percolation, need to be kept in mind. First,
unlike in the connectivity case, a site can belong to more than one cluster
(it then serves as a pivot joint between the clusters sharing this site). On
the other hand, a bond in 2D always belongs to just one cluster (not so in 3D,
when it can serve as a hinge around which several clusters can rotate). For
this reason, cluster decomposition in 2D is best expressed in terms of
bonds and not sites. While this is not generally so, in the case of
self-organized networks in the floppy or intermediate phase, the conversion
between cluster sizes expressed in terms of bonds or sites is easy: since
there is no stress, there is also no redundancy, and every rigid cluster of
$n$ sites contains exactly $2n-3$ bonds.

We start with properties of the percolating cluster.

\subsection{Definition of a percolating cluster}

Normally, when the probability of percolation is either zero or one in the
thermodynamic limit, the exact definition of percolation does not matter. In our
case, since the probability of finding a percolating cluster increases linearly with
average coordination in the intermediate phase, it is not clear that we can be so
cavalier.

For example, the probability that a cluster that percolates {\it in only one
direction} (no matter which one) seems to be non-zero in the intermediate
phase, although it is very low, around 0.1 or less. This means that the
results will differ slightly if we define percolation using only one direction
or both. To avoid confusion, we choose here to call percolating those networks
in which percolation occurs in both directions, and non-percolating those
networks in which there is no percolation in either direction. Networks with
percolation in just one direction are ignored whenever we separate our results
into those for percolating networks and those for non-percolating ones, but
such networks are taken into account when this separation is not done.

\subsection{Size of the percolating cluster}

\begin{figure}
\begin{center}
\includegraphics[width=2.7in]{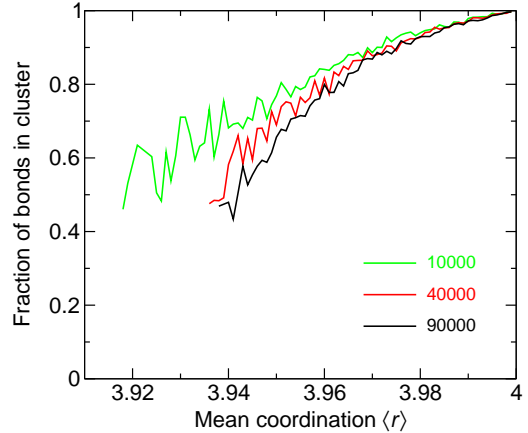}
\end{center}
\caption{The fraction of bonds belonging to the percolating rigid cluster among
all bonds in the network,
averaged over all percolating self-organized networks, for different network
sizes indicated in the figure. All sizes here and in other figures in the paper
are given in terms of sites.}
\label{fig:perc_clst}
\end{figure}

While we have already studied the probability that a percolating cluster occurs in
the intermediate phase, its size had not been characterized.
Figure~\ref{fig:perc_clst} shows the average number of bonds in the percolating
cluster as a fraction of bonds {\it actually present} in the network. The averaging is
done over all networks in which percolation occurs. A remarkable feature is that even
at the lowest $\ra$ at which percolation is still (rarely) observed, the size of the
percolating cluster is well above zero. The smallest cluster size observed at the
onset of the intermediate phase is around 40\%. This behavior is different from that
for both connectivity and rigidity percolation on regular lattices in the random (non-self-organized)
case, where the size of the percolating cluster, considered a good order parameter,
grows from zero at the transition, as expected in a second-order transition. This
result in the self-organized case is reminiscent of the {\it first-order} rigidity
transition, such as that observed on Bethe lattices.~\cite{dux97, dux99, travthorpe}.
However, our other results, as discussed below, do not support this analogy.

\begin{figure}
\begin{center}
\includegraphics[width=2.7in]{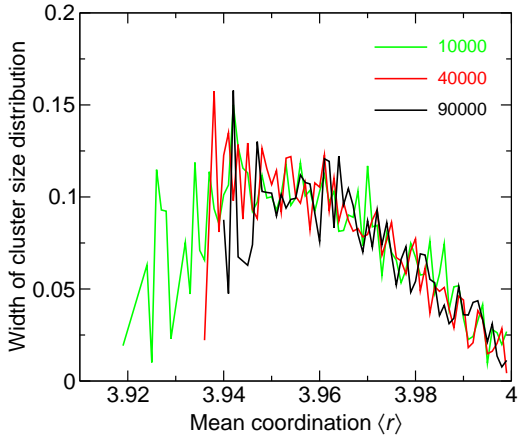}
\end{center}
\caption{The standard deviation of the fraction of bonds in the percolating
cluster for different network sizes, for self-organized networks.}
\label{fig:perc_width}
\end{figure}

Figure~\ref{fig:perc_clst} presents {\it average} sizes of percolating clusters.
Given that the very existence of the percolating cluster is uncertain in the
intermediate phase (since only some networks are percolating), it is reasonable
to ask about the variation of the percolating cluster size. The
quantity we look at is the standard deviation, or width, of the fraction of
bonds in the percolating cluster calculated as $\sqrt{(\langle F^2\rangle
-\langle F\rangle ^2)n/(n-1)}$, where $F$ is the fraction of bonds in the
percolating cluster, $\langle\ldots\rangle$ denotes the average over percolating
networks and $n$ is the number of percolating networks.  In ordinary
percolation, this width, of course, tends to zero as the network size
grows; the percolating cluster size is a self-averaging quantity.
Figure~\ref{fig:perc_width} shows that this is not so in our case. The width is
above zero and is essentially size-independent. This is yet another difference
from non-self-organized percolation (including that on Bethe lattices). Note
at the same time that the width of the distribution of percolating cluster
sizes is much smaller than the average size. In other words, an overwhelming
majority of networks have either a big percolating cluster or no percolating
cluster at all --- there are few (if any) ``intermediate cases'' with small
percolating clusters.

\subsection{Sizes of non-percolating rigid clusters}
 
To further characterize the intermediate phase, it is useful to look at the
distribution of rigid cluster sizes. In case of a second-order phase
transition, the correlation radius is finite away from the transition. As a result, at a
certain cluster size there is a crossover from a power-law behavior to an
exponential behavior. The correlation radius
diverges as the transition is approached from either side, and so the crossover
moves towards bigger sizes as the
transition is approached, and exactly at the transition, the power law
persists indefinitely. Since the divergence of the correlation radius (or the
crossover point) is governed by the same critical exponent on both sides of
the transition, it is expected that for two values of $\ra$ at the same
distance from the transition but on opposite sides, the crossover points will
be the same by order of magnitude, if the transition is second-order. In
case of a first-order transition, the correlation radius does not diverge at
the transition and there is always a crossover to the exponential behavior.

\begin{figure}
\begin{center}
\includegraphics[width=2.7in]{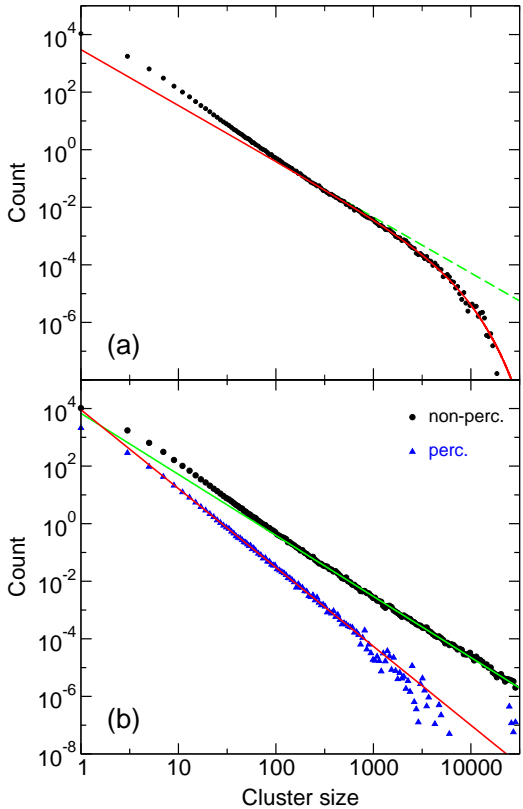}
\end{center}
\caption{The distribution of sizes of non-percolating rigid clusters in
non-percolating self-organized networks at $\ra=3.92$ [panel (a)] and in both
percolating and non-percolating self-organized networks at $\ra=3.97$
[panel (b)]. The details of the simulation and the fits (lines) are given in the
text.}
\label{fig:clst_dist}
\end{figure}

Figure~\ref{fig:clst_dist} shows the distribution of rigid cluster sizes at the
coordinations of $\ra=3.92$ (below the transition) and $\ra=3.97$ (above the
transition). The points are chosen at about the same distance from the rigidity
transition, far enough from it to make sure that the correlation radius is not too
big (if the transition is second-order) yet not too far to ensure that the points are
still within the critical region. All results are obtained by running 200 independent
simulations on networks of 40000 sites, each starting from an already equilibrated
network and continuing for 1000 additional equilibration steps. After each
equilibration step, the distribution of non-percolating cluster sizes is obtained.
While networks one equilibration step apart cannot be considered truly independent,
inserting or removing a single bond often changes the rigidity of the network very
significantly, justifying inclusion of
the data obtained at every step. For $\ra=3.97$, the distributions for
percolating and non-percolating networks are plotted separately (at $\ra=3.92$,
obviously, there are very few percolating networks and we plot the result for
non-percolating networks only). To decrease the noise in the tail, all clusters of a
given class (i.e., percolating and non-percolating) are binned using a logarithmic
scale.

Below the rigidity transition, in the floppy phase [$\ra=3.92$;
Fig.~\ref{fig:clst_dist},(a)], there is a clear crossover between the power-law
and exponential behaviors. The data are fitted using the product of a
power law and an exponential:
\begin{equation}
f_1(x)=C_1 x^{-\alpha_1}\exp(-x/x_0).\label{eq:fit1}
\end{equation}
We use the data between $10^2$ and $\approx 10^{4.23}$, i.e., dropping just a
few data points at the tail, where noise and finite-size
effects are significant, and omitting a region at the low end,
as there are big deviations from the behavior described by Eq.~\ref{eq:fit1},
probably due to discretization effects from the lattice. The best fit is
obtained with $C_1=3000$, $\alpha_1=1.94$ and $x_0=3900$. The dashed line
is a power law with the same $C_1$ and $\alpha_1$, but without the exponential
factor.

Above the transition [$\ra=3.97$; Fig.~\ref{fig:clst_dist},(b)], however, the
power-law behavior persists at least for non-percolating networks, with no hint
of the exponential tail, even for the largest sizes for which the data are
available (around 30000, well above the crossover observed for $\ra=3.92$ around
$x_0=3900$). For percolating nets, there is some deviation from the power law near
the end, but it is likely due to finite-size effects (there is a percolating
cluster taking up most of the network, so only relatively small non-percolating
clusters are possible). To fit the data, we use pure power-law functions:
\begin{equation}
f_{2\{n,p\}}=C_{2\{n,p\}} x^{-\alpha_{2\{n,p\}}},\label{eq:fit2}
\end{equation}
where subscripts $\{n,p\}$ refer to non-percolating and percolating networks,
respectively. We use data above $10^2$ in the non-percolating case and from $10^{1.8}$
to $10^{2.8}$ in the percolating case. The values of the parameters providing the best
fits are: $C_{2n}=6800$, $\alpha_{2n}=2.12$, $C_{2p}=9100$, $\alpha_{2p}=2.74$. While
the difference between $\alpha_1$ and $\alpha_{2n}$ is probably due to finite size
and sampling effects, the difference between $\alpha_{2n}$ and $\alpha_{2p}$ suggests
that these two quantities are different, an unusual behavior.

Our results indicate that the power-law distribution of cluster sizes is
observed in the whole intermediate phase, rather than at a single point (as
would be the case without self-organization). In effect,
the self-organization, which minimizes the stress in the network,
\textit{maintains the system in a critical state throughout the intermediate
phase}. Similarly, unlike in the case of the critical point of
random percolation, the cluster size critical exponent appears to be different
in the non-percolating and percolating cases. 

Based on the results of this section, we see that the rigidity phase
transition in our model of self-organized percolation is very different from
both first- and second-order phase transitions. The minimization of the stress
appears to act similarly to an external driving force, leading to a self-organized
critical phase in a thermodynamically equilibrated system. 

\section{Response to a local perturbation}
\label{sec:pertur}

Given that both percolating and non-percolating networks coexist in the
intermediate phase, it is interesting to investigate the relation between the
two classes. As we show in this section, a local perturbation involving the
addition or removal of even a single bond is enough to affect rigidity of huge
regions of the network and often converts a non-percolating network into a
percolating one and vice versa.

This behavior is not observed in regular random networks: since the
probability of percolation in these systems is always either zero or one in
the thermodynamic limit, a single bond can only change the percolation
property right at the transition. Since a percolating network at that point is
fractal and involves only an infinitesimal fraction of bonds, if an infinite
cluster is created or destroyed, this can only involve an infinitesimal
fraction of bonds and sites; away from the transition, the size of the
affected region is always finite and then the fraction is obviously
infinitesimally small.

A few general comments about the consequences of addition or removal of a
single bond are in order. First of all, the removal of a single bond can only
break up the cluster to which this bond belongs; other clusters are not
affected. This is because in 2D, all rigid clusters are always rigid by
themselves, i.e., they remain rigid when taken in isolation from the rest of
the network. Conversely, the addition of a bond can merge several rigid clusters
into one, but will not affect the clusters outside the resulting cluster. Note
that even though the self-organized networks are by definition stress-free, a
newly inserted bond can be redundant and introduce stress (always confined to
the cluster in which it is inserted); in this case the
configuration of rigid clusters is not affected, but the created stressed
region may still be macroscopic and percolate. While we will briefly consider
this situation at the end of this section, we mostly concentrate on the case
when the inserted bond is non-redundant and thus affects the rigidity of the
network but does not create stress. For brevity, we will call the places where
insertion of a bond does not create stress, as well as a bond inserted in such
a place, {\it allowed}. Note that if we add an allowed bond creating a certain
rigid cluster and then remove this bond and insert another allowed bond both
ends of which are in the region coinciding with the cluster created by the
first bond, exactly the same cluster will be created: indeed, the count of
constraints within the region will be the same ($2n-3$ for a region of $n$
sites), regardless of where in this region the bond is inserted.

\subsection{Conversion between percolating and non-percolating networks}

We first look at the conversion between non-percolating and percolating
networks as a result of bond addition and removal. There are several related
questions here.

\subsubsection{Addition of a bond in an allowed position}

\begin{figure}
\begin{center}
\includegraphics[width=2.7in]{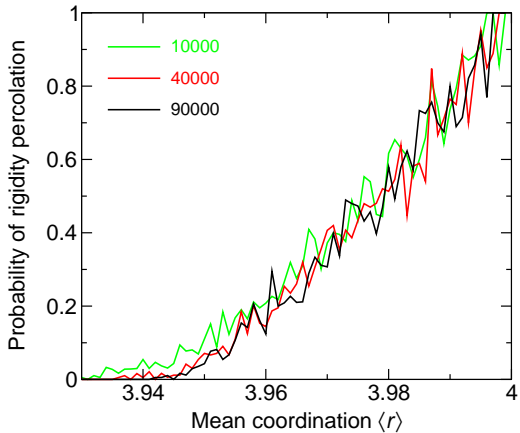}
\end{center}
\caption{The probability that a bond added to a non-percolating self-organized
network at a random allowed place makes this network percolating.}
\label{fig:ins1}
\end{figure}

In Fig.~\ref{fig:ins1}, we plot the probability that inserting a bond at a
\textit{randomly chosen} allowed place makes a non-percolating network
percolating. In the
floppy phase (below $\ra\approx 3.945$), this probability tends to zero in the
thermodynamics limit; however, it is non-zero everywhere in the intermediate
phase. This probability is close to zero just above the rigidity transition
and approaches one close to the upper boundary of the intermediate phase at
$\ra=4$.

\begin{figure}
\begin{center}
\includegraphics[width=2.7in]{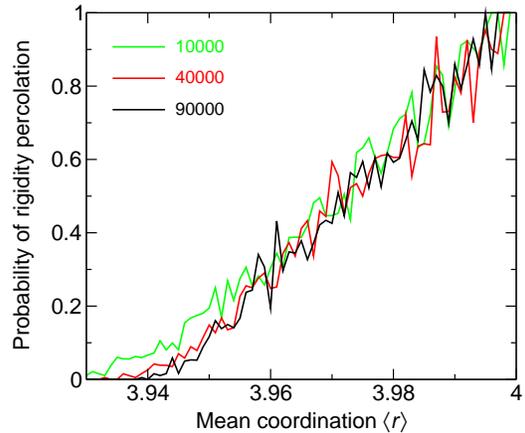}
\end{center}
\caption{The fraction of non-percolating self-organized networks that can become
percolating upon a single bond addition.}
\label{fig:ins2}
\end{figure}

We can also ask about the frequency of networks in which conversion from
non-percolating to percolating upon a single bond addition is possible.
Note that because of the above-mentioned property that insertion of a bond
into a region made rigid by another bond rigidifies exactly the same region,
we do not need to try each and every allowed bond, which speeds up the
simulation enormously. 
Fig.~\ref{fig:ins2} plots the fraction of non-percolating networks that can
become percolating with a bond placed judiciously in an allowed position. Note
that this is different from the previous figure: instead of choosing a place
to insert a bond at random, we now make the best effort to cause percolation
with a single bond addition, if at all possible. As is expected, this
quantity is higher than that plotted in Fig.~\ref{fig:ins1}, but the
difference is  small; over most of the intermediate phase,
whenever there are \textit{any} allowed positions where bond insertion creates
a percolating cluster, {\it most} allowed positions will do.

It is also interesting to note that the quantity in Fig.~\ref{fig:ins2} is
very close to linear in the intermediate phase, and is probably exactly
linear, just like the probability of percolation without any bond insertions.
In the Appendix, we explain why these two quantities are equal.

\begin{figure}
\begin{center}
\includegraphics[width=2.7in]{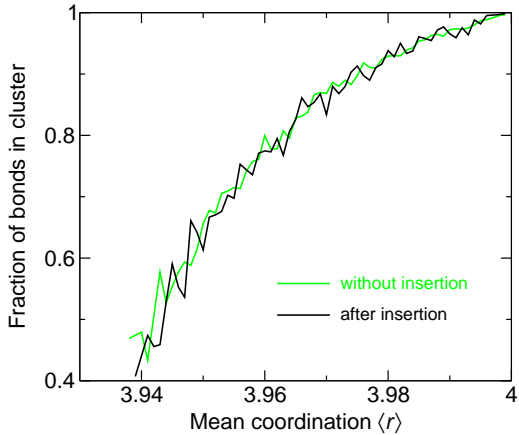}
\end{center}
\caption{The comparison between the average size of the percolating cluster in
originally percolating self-organized networks (``without insertion'') and in
originally non-percolating networks that become percolating after bond insertion
(``after insertion''). The latter quantity is calculated as the average over
all networks that can become percolating, by using ``judicious placement'' of
a bond, as described in the text.}
\label{fig:clst_compare}
\end{figure}

Finally, in Fig.~\ref{fig:clst_compare}, the average size of the percolating cluster
arising after bond insertion is compared to the average size of the percolating
cluster in those cases when it exists even without bond insertion (i.e., the quantity
in Fig.~\ref{fig:perc_clst}). These values are identical: the size of the percolating
cluster emerging after inserting a bond in a non-percolating network is the same as
in originally percolating networks. This suggests that a percolating network that
arises after bond insertion is a typical percolating network, just like those
networks that percolate without insertion. In the next subsection, we present
more evidence in favor of this. Also see a more detailed discussion of this and
a possible caveat in the Appendix to this paper.

Given the significant probability of conversion of non-percolating
networks to percolating ones (see Fig.~\ref{fig:ins2}), we can
hypothesize that {\it all} non-percolating networks can become
percolating after a {\it finite} number of bond insertions; moreover, the
average size of the percolating cluster after the minimal number of insertions
needed to create it is again the same as the average size of
the percolating cluster in networks that are percolating without bond
insertions. An even stronger hypothesis is that after each new bond addition,
the fraction of so far non-percolating networks becoming percolating is
still the same linear function of $\ra$ as in Fig.~\ref{fig:ins2}. These
hypotheses need to be tested in the future.

\subsubsection{Removing a bond from the network}

\begin{figure}
\begin{center}
\includegraphics[width=2.7in]{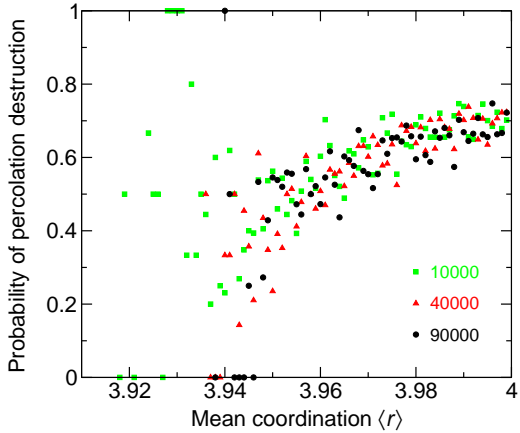}
\end{center}
\caption{The probability that removal of a randomly chosen bond from a
percolating self-organized network destroys percolation.}
\label{fig:rem}
\end{figure}

\begin{figure}
\begin{center}
\includegraphics[width=2.7in]{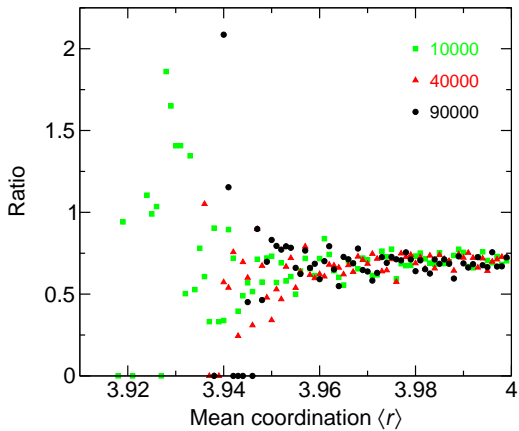}
\end{center}
\caption{The probability that removal of a random bond destroys percolation
(the quantity in Fig.~\ref{fig:rem}) divided by the average fraction of bonds
in the percolating cluster (the quantity in Fig.~\ref{fig:perc_clst}). This
serves as an estimate of the probability that a bond chosen at random among
those belonging to the percolating cluster destroys percolation.}
\label{fig:redbonds}
\end{figure}

Likewise, we can consider the probability that removing a randomly chosen bond from a
percolating network breaks the percolating cluster. This quantity is shown in
Fig.~\ref{fig:rem}. Just like for bond removal, this probability is non-zero everywhere
in the intermediate phase. Interestingly, even at the upper boundary ($\ra\to 4$),
where almost all networks are percolating, it is still very easy to break
percolation and thus create a non-percolating network. In fact, the
probability is the highest in this limit. Of course, this
can be attributed to the fact that the percolating cluster is itself the biggest at
this point (taking up the whole network), so there is a greater chance than elsewhere
to select a bond that belongs to it (which is, of course, a necessary condition of
its destruction). This effect can be factored out by dividing the quantity in
Fig.~\ref{fig:rem} by the average fraction of bonds in the percolating cluster. This
will then give the probability of the destruction of the percolating cluster, {\it
given that} the removed bond belongs to this cluster. This is plotted in
Fig.~\ref{fig:redbonds}. It is seen that across the whole intermediate phase, this
quantity is nearly constant at around 70\%. In other words, everywhere in the
intermediate phase, the percolating cluster on average contains about 70\% of bonds
such that removal of any one of them will destroy percolation. This is true even
close to $\ra=4$. This is, of course, just the average; one could ask if
for some networks this quantity is zero (similar to how in the case of
bond insertion, not all non-percolating networks can be made percolating by a single
bond addition, as shown in Fig.~\ref{fig:ins2}). To check if this is the case, we
remove (and then reinsert) up to 10 bonds one by one (all chosen within the
percolating cluster) and see if the percolating cluster ever gets destroyed. This
happened for 5398 out of 5429 90000-site networks, or about 99.4\%, including 198 out
of 200 networks (99\%) at $\ra=3.999$ (the highest mean coordination in our
simulations). Since these fractions are so close to 100\%, we can hypothesize that,
unlike in the case of bond insertion, in fact, {\it all} percolating networks have
some bonds whose removal destroys percolation.

To conclude this subsection, our results for the change in the percolation state
of the network upon single bond addition or removal again confirm that the
system remains critical in the whole intermediate phase.

\subsection{Change of rigidity: ``rigid'' and ``floppy'' bonds}

In the previous subsection, we have found that in many cases insertion or
removal of a single bond can change the percolation status of the network and
affect the rigidity of its significant part. Since we were dealing with
percolating clusters only, we could not study the effect of bond insertion or
removal in those cases when the percolation status does not change. To do this,
we introduce the concept of ``rigid'' and ``floppy'' bonds.

\begin{figure}
\begin{center}
\includegraphics[width=2.7in]{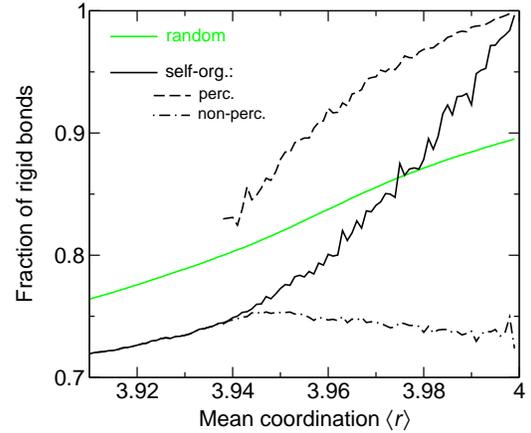}
\end{center}
\caption{The average fraction of rigid bonds in the network in the random and
self-organized cases (in the latter case, also separately for percolating and
non-percolating networks). All simulations are for networks of 90000 sites, the
overall averages are over 200 networks, the averages restricted to percolating
and non-percolating networks are over those of the 200 self-organized networks
that are respectively percolating and non-percolating.}
\label{fig:rigbonds}
\end{figure}

Below the rigidity transition, only small rigid clusters are present in the
network. Interestingly, a significant fraction of them consist of just a
single bond. The number of bonds
belonging to these single-bond clusters decreases as the rigidity transition
is approached and eventually crossed, although they are still encountered well
above the transition in floppy pockets of the network. Since these bonds are
associated with the floppy phase and floppy regions of the network in the
rigid phase, we call such bonds {\it floppy}. All other bonds (i.e., those
belonging to clusters consisting of more than one bond, or, in other words,
rigid with respect to some other bonds) are called {\it rigid}. The fraction
of rigid bonds as a function of $\ra$ is plotted in Fig.~\ref{fig:rigbonds}
for both random and self-organized networks. As expected, it grows in both
cases with increasing $\ra$, but in the self-organized case it reaches 1 at
$\ra=4$ (when the network becomes fully rigid), which in the random case does
not happen until the full coordination at $\ra=6$. For the self-organized case,
the averages over just percolating networks and over just non-percolating
networks are also shown in Fig.~\ref{fig:rigbonds}. Interestingly, the average
for non-percolating networks remains nearly constant over the whole intermediate
phase at about 75\%, even as $\ra\to 4$. In this limit, the number of floppy
modes per site tends to zero in all networks; yet, as these results show, in
those few networks that still do not percolate, many bonds (about 25\%) are
still in single-bond clusters. 

\begin{figure}
\begin{center}
\includegraphics[width=2.7in]{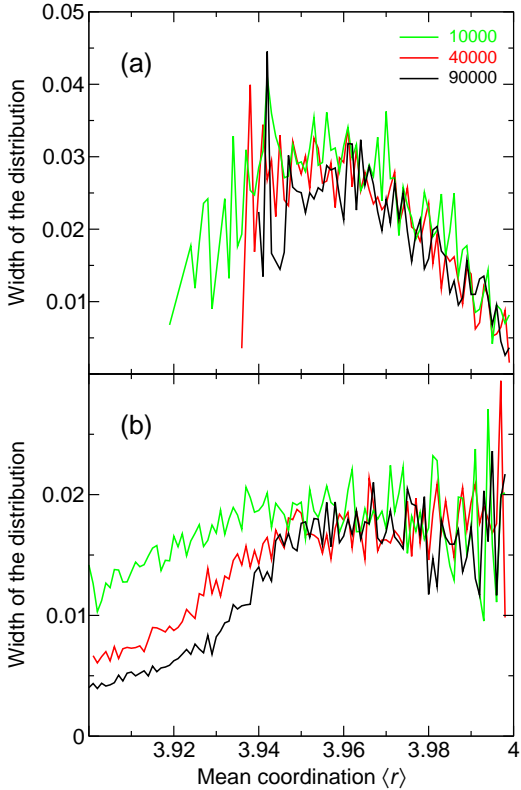}
\end{center}
\caption{The standard deviation of the distribution of fractions of rigid
bonds in percolating [panel(a)] and non-percolating [panel (b)] self-organized
networks for different network sizes.}
\label{fig:rigbonds_width}
\end{figure}

Just as in the previous section we considered the width of the distribution of
percolating cluster sizes, it is interesting to look at the width of the
distribution of fractions of rigid bonds. Of course, since the averages are
very different for percolating and non-percolating networks, it makes sense to
separate these two classes. The results are shown in
Fig.~\ref{fig:rigbonds_width}. Note that in the intermediate phase, the width
is size-independent and so likely remains finite in the thermodynamic limit,
for both percolating and non-percolating networks, just as we have seen for the
percolating cluster size (see Fig.~\ref{fig:perc_width}). But in the floppy
phase (where, of course, only non-percolating networks are present), the width
clearly decreases fast with size (it is roughly inversely proportional to the
square root of the network size). Thus the number of rigid bonds is a
self-averaging quantity in the floppy phase but not in the intermediate phase,
even when percolating and non-percolating networks are considered separately.
At the same time, the widths for both percolating and non-percolating cases are
much smaller that the difference between these two cases, so based on the
count of rigid bonds, these two classes are clearly distinct.

We now look at the influence of insertion or removal of a bond on the number
of rigid bonds and their spatial distribution.

\begin{figure}
\begin{center}
\includegraphics[width=2.7in]{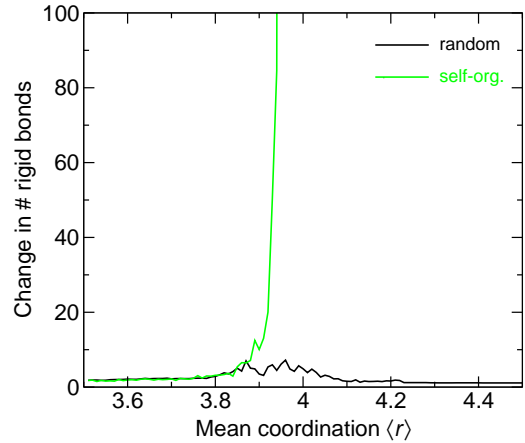}
\end{center}
\caption{The average change in the number of rigid bonds upon addition of a
single bond, for random and self-organized networks of 90000 sites.}
\label{fig:bonds_flop2rig2}
\end{figure}

In Fig.~\ref{fig:bonds_flop2rig2}, we show the average change in the number of rigid
bonds upon {\it insertion} of a single bond, in both the random and the
self-organized cases. Note that in the random case, this change is very small on
average, around 10 bonds or less. On the other hand, in the self-organized case this
quantity diverges very fast when the rigidity transition is approached. This is not
surprising: we have seen that infinite percolating clusters can easily form and break
in this case. For this reason, it makes sense to look at the {\it fraction} of bonds
that undergo the change.

\begin{figure*}
\begin{center}
\includegraphics[width=6.0in]{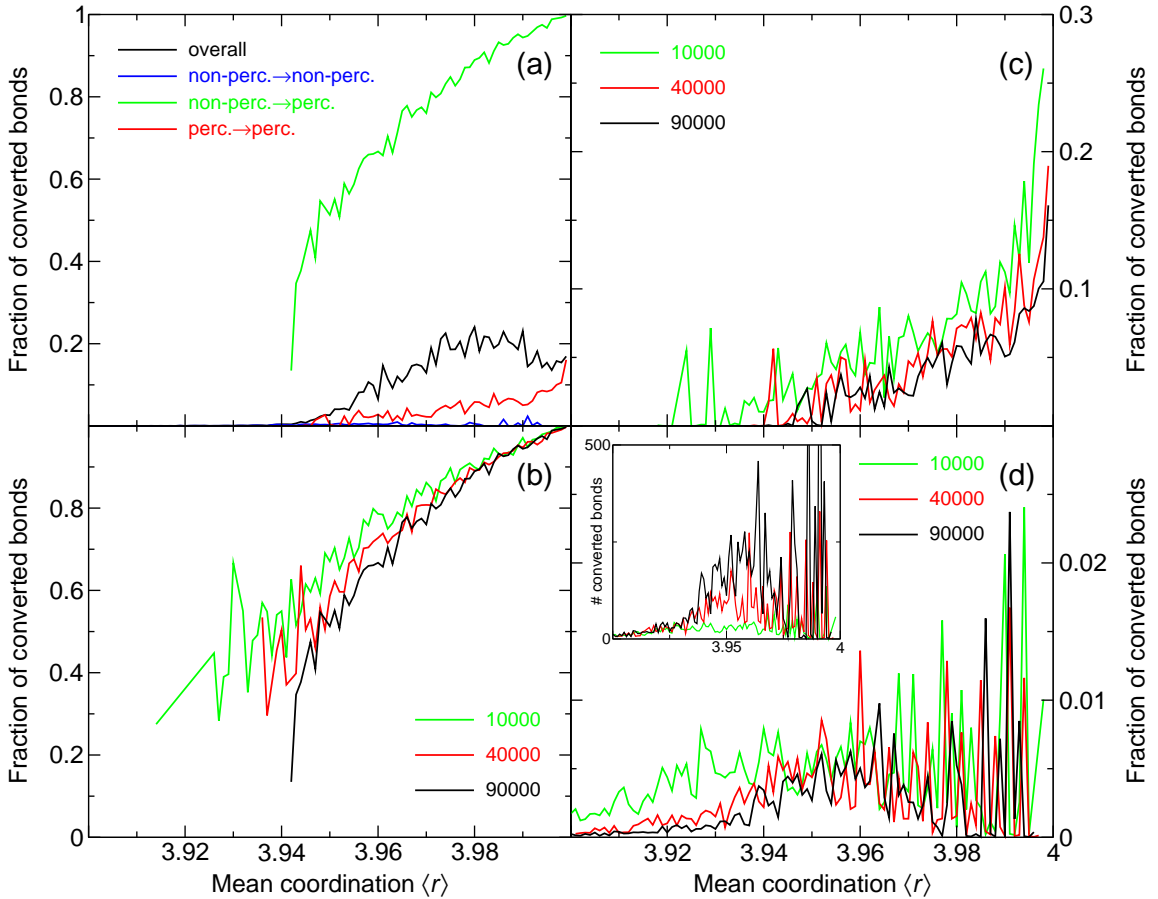}
\end{center}
\caption{The average fraction of floppy bonds converting to rigid upon addition
of a single bond to a self-organized network. Panel (a) shows the overall
average, as well as partial averages for the cases when the network converts
from non-percolating to percolating, remains non-percolating and remains
percolating, for networks of 90000 sites. Panels (b), (c), and (d) show the
partial averages in the non-percolating $\to$ percolating,
percolating $\to$ percolating, and non-percolating $\to$ non-percolating cases,
respectively, for three different network sizes. The inset in panel (d) shows
the average {\it number} of converted bonds in the non-percolating $\to$
non-percolating case.}
\label{fig:frac_flop2rig}
\end{figure*}

In Fig.~\ref{fig:frac_flop2rig}, we show the fraction of bonds converted from floppy
to rigid when a bond is added to the network at a randomly-chosen allowed
position, among all {\it floppy bonds} in the network. In panel (a), we plot the
overall average, as well as partial averages restricted to those cases when
the network goes from non-percolating to percolating, remains percolating
and remains non-percolating (the curve for the latter case is barely above the
baseline). In panels (b), (c), and (d), we show these
partial averages for several different sizes. Based on what we have already
learned for networks switching from non-percolating to percolating, it is
quite natural that in this case a significant fraction of
floppy bonds become rigid; this fraction approaches 1 when $\ra\to 4$, which is
again expected, since in this limit the percolating cluster takes up the whole
network (see Fig.~\ref{fig:clst_compare}). The other two cases are more
interesting. In the floppy phase, only the non-percolating $\to$
non-percolating situation is possible, and as we see from panel (d), in this
region of the phase diagram the average fraction of bonds converted from floppy
to rigid decreases fast with size. In the inset, we plot the average {\it number}
(rather than fraction) of converted bonds, and we see that this number is
size-independent. So in the floppy phase, a {\it finite} number of bonds gets
converted. In the intermediate phase, the situation is different. Looking first
at the percolating $\to$ percolating situation [panel (c)], the fraction of
converted bonds depends only weakly on the network size, and it is possible that
this quantity goes to a constant in the thermodynamic limit. Even if, in fact,
this fraction decreases to zero as the size $N$ goes to infinity, it is clear
that the decrease is much slower than $\propto 1/N$, and thus the mean
{\it number} of converted bonds diverges when $N\to\infty$. In the
non-percolating $\to$ non-percolating case, the quality of the data is lower
(there is very much noise due to a very large variation in the number of
converted bonds), but still it is clear that the decrease with $N$ (if present)
is certainly much slower than $\propto 1/N$.

The fact that for the percolating $\to$ percolating and non-percolating $\to$
non-percolating cases the average number of converted bonds diverges in the
thermodynamic limit in the intermediate phase means that even when the
percolation status does not
change, the region of the network whose rigidity is affected are
macroscopic {\it at least in some cases}. This is again consistent with the
criticality of the intermediate phase. The distribution of the sizes of affected
regions (or numbers of converted bonds) is likely power-law, which needs to
be tested in the future.

\begin{figure}
\begin{center}
\subfigure{
\includegraphics[width=2.7in]{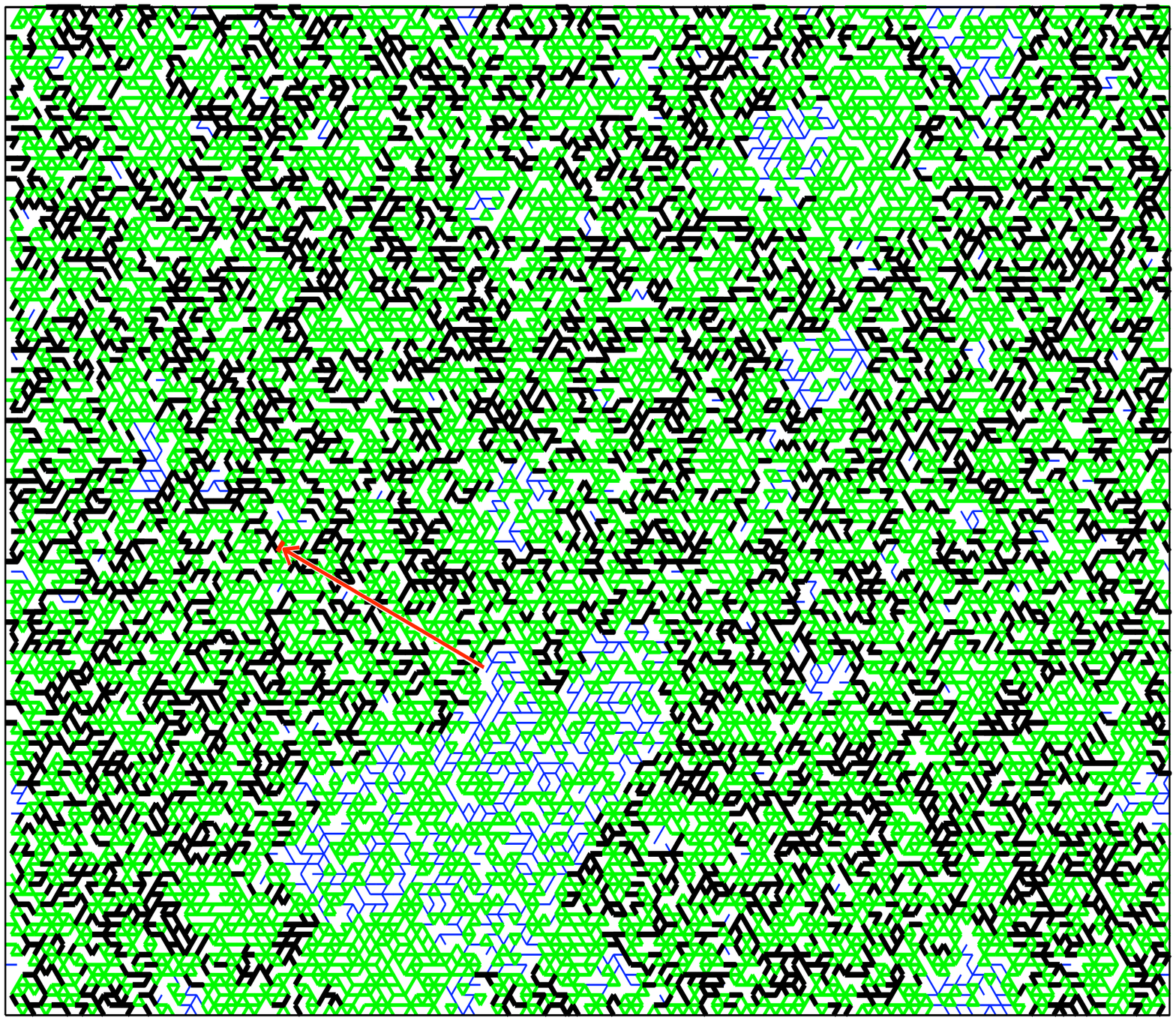}
}
\subfigure{
\includegraphics[width=2.7in]{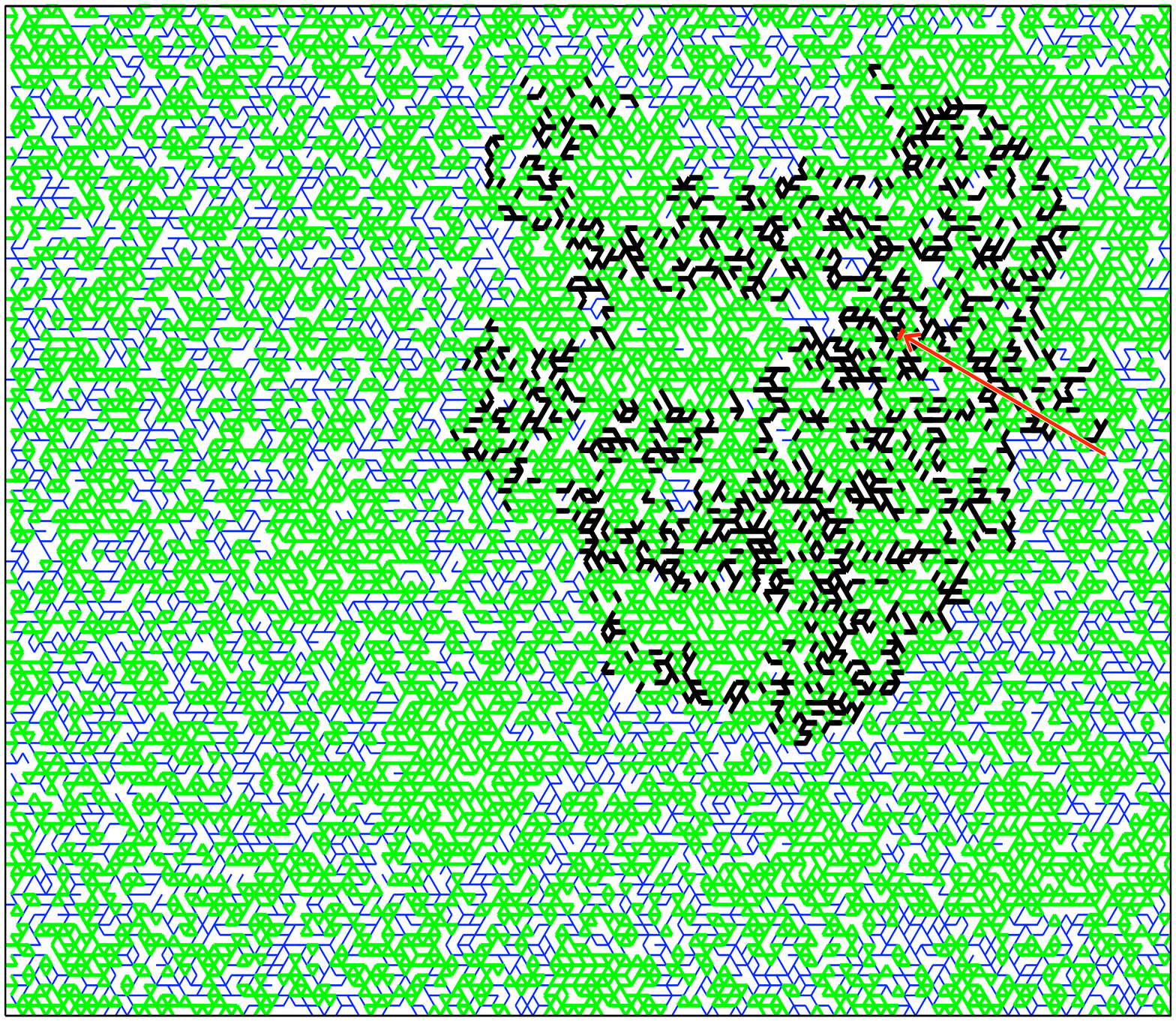}
}
\end{center}
\caption{Two examples of rigidification of self-organized networks upon addition
of a bond. In the top panel, the network switches from non-percolating to
percolating. In the bottom panel, it remains non-percolating. In both cases, the
added bond is red and pointed with an arrow, thick green
bonds are those that are originally rigid, thin blue bonds remain floppy,
thick black bonds switch from floppy to rigid. Both networks contain 10000 sites.}
\label{fig:net}
\end{figure}

As an illustration of effects of bond insertions we show two examples in
Fig.~\ref{fig:net}. The upper panel shows an example for the case when
the network switches from non-percolating to percolating after a bond is added.
The lower panel shows an event where the network remains non-percolating, but
large-scale rigidification still occurs without percolation.
In both cases, the added bond is red and pointed with an arrow, thick green
bonds are those that are originally rigid (and, of course, remain rigid after
bond addition), thin blue bonds are originally floppy and remain floppy,
finally, thick black bonds are of most interest: these are the ones that switch
from floppy to rigid. In the first
(non-percolating $\to$ percolating) case, converted bonds are spread throughout
the network. Many bigger rigid clusters
separated by floppy ``interfaces'' merge into one percolating rigid cluster; in essence, the
figure illustrates the rigidification of these interfaces. In the second
(non-percolating $\to$ non-percolating) example, the affected region is still
large, but non-percolating. This
example is larger than average, as can be deduced from
Fig.~\ref{fig:frac_flop2rig}, panel (d), but it is not a rare event.

\begin{figure*}
\begin{center}
\includegraphics[width=6.0in]{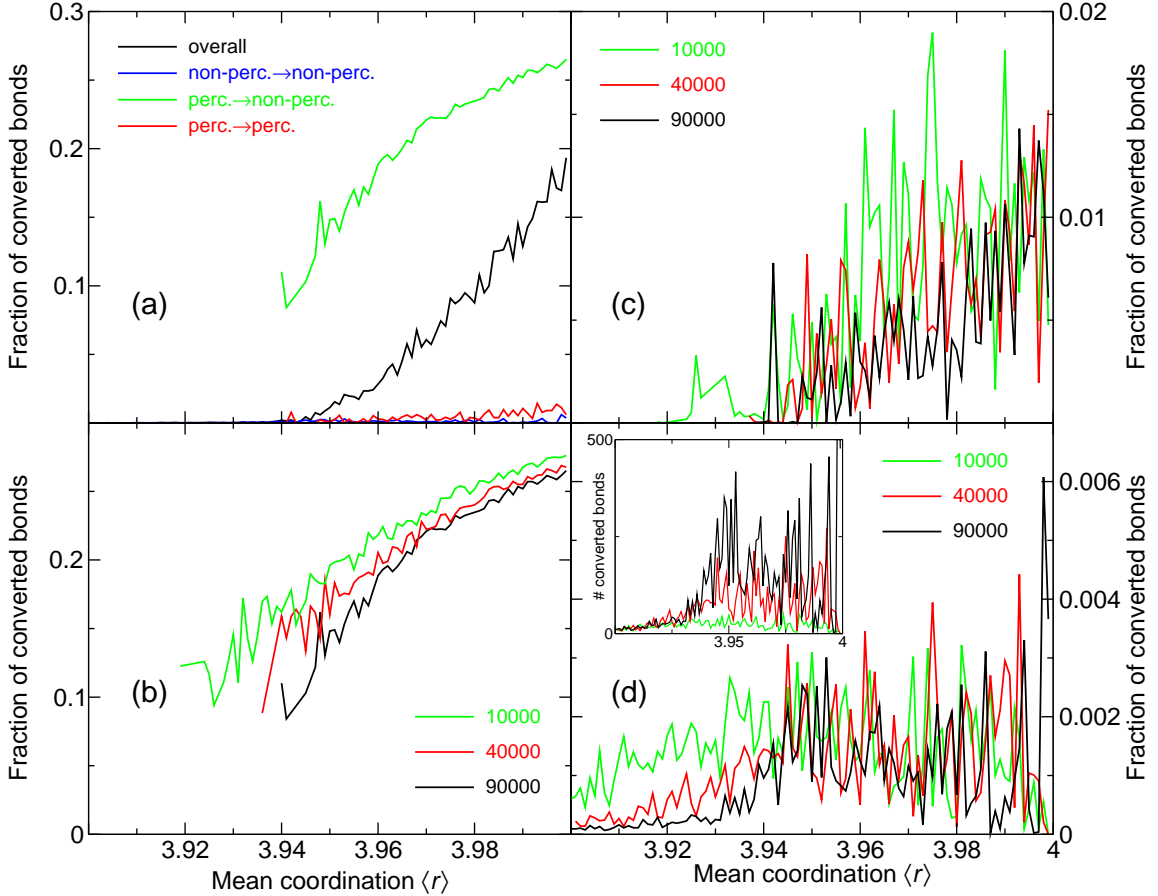}
\end{center}
\caption{The average fraction of rigid bonds converting to floppy upon removal
of a single bond from a self-organized network. Panel (a) shows the overall
average, as well as partial averages for the cases when the network converts
from percolating to non-percolating, remains non-percolating and remains
percolating, for networks of 90000 sites. Panels (b), (c), and (d) show the
partial averages in the percolating $\to$ non-percolating,
percolating $\to$ percolating, and non-percolating $\to$ non-percolating cases,
respectively, for three different network sizes. The inset in panel (d) shows
the average {\it number} of converted bonds in the non-percolating $\to$
non-percolating case.}
\label{fig:frac_removal}
\end{figure*}

We find overall similar behavior upon {\it removal} of a bond.
The results for the fraction of rigid bonds converting to floppy are in
Fig.~\ref{fig:frac_removal}. Similarly to Fig.~\ref{fig:frac_flop2rig}, in
panel (a) we have the overall average and partial averages for
percolating $\to$ non-percolating, percolating $\to$ percolating and
non-percolating $\to$ non-percolating cases, for a single network size
(90000 sites). In panels (b), (c), and (d), we have the same partial averages,
but for three different sizes. Again, the conclusions are similar to the
case of bond insertion: in the percolating $\to$ non-percolating case, when
the percolation status of the network changes [panel (b)], the fraction of bonds
that switch from rigid to floppy is expectedly high. It is much lower in the
other two cases, when the percolation status does not change, but still, just
as for bond addition, while in the floppy phase the fraction of converting bonds
falls rapidly with increasing size and, as the inset of panel (d) shows, the
average {\it number} of converting bonds remains constant, in the intermediate
phase, again, the dependence of the fraction of converting bonds on the size is
very slow and the number of converting bonds diverges in the thermodynamic
limit --- thus again, macroscopic regions of the network can be involved.

\begin{figure}
\begin{center}
\includegraphics[width=2.7in]{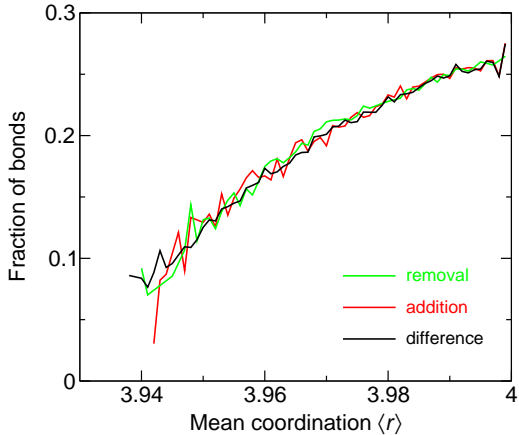}
\end{center}
\caption{The average fractions of bonds:
converting from rigid to floppy when a bond is removed and the network switches
from percolating to non-percolating (green); converting from
floppy to rigid when a bond is inserted and the network switches from
non-percolating to percolating (red).
These are compared to the difference between the average fractions of rigid
bonds in the percolating and non-percolating cases. All quantities are
for networks of 90000 sites.}
\label{fig:backnforth}
\end{figure}

We finish this subsection with an observation: there exists
a symmetry in bond conversions upon addition and
upon removal. Namely, the average number of bonds converting upon bond
addition in the case when the network
transforms from non-percolating to percolating is the same as the average number
of bonds converting upon bond removal when the network transforms from
percolating to non-percolating. This is illustrated in
Fig.~\ref{fig:backnforth}. In this figure, we plot the ratios of the
above-mentioned numbers and the total number of bonds in the network. We find
that these two quantities coincide. Note that these quantities are different
from those plotted in
Figs.~\ref{fig:frac_flop2rig} and \ref{fig:frac_removal}: in these figures,
the numbers of converted bonds were divided by the number of floppy bonds and
the number of rigid bonds, respectively, and not by the total number of bonds.
The equality is easy to understand, if we recall that we have already seen
some evidence (see Fig.~\ref{fig:clst_compare}) that networks that become
percolating after bond insertion are, in fact, typical percolating networks,
just like those that are originally percolating. By extension, we can assume
that networks that become non-percolating after bond removal are also typical
non-percolating networks. If so, then the average change in the number of
rigid bonds should in both cases be the same as the difference in the
average number of rigid bonds between percolating and non-percolating networks.
Indeed, in Fig.~\ref{fig:backnforth} we also plot the difference between the
average fractions of rigid bonds in the non-percolating and percolating cases
(i.e., between the dashed and the dot-dashed lines in Fig.~\ref{fig:rigbonds});
it is seen that this quantity coincides with the other two.

\subsection{Stress propagation}

In the previous subsections, we have looked at cases when an ``allowed'' bond insertion
is done; in other words, the bond is inserted in one of those places where it does
not create stress. We now look at the opposite situation, i.e., we analyze the
results of inserting a bond in one of ``disallowed'' places. In this case, the
inserted bond is {\it redundant}, which means that its insertion does not change the
configuration of rigid clusters, but some bonds (including the inserted one) become
stressed. We are interested in the emerging stressed region and in particular,
whether it percolates or not.

\begin{figure}
\begin{center}
\includegraphics[width=2.7in]{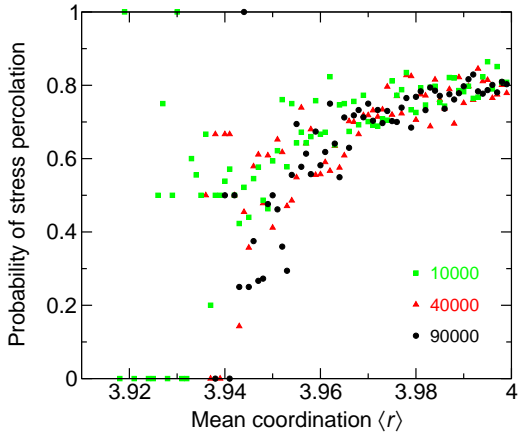}
\end{center}
\caption{The probability that a percolating stressed region forms when a bond is
inserted at a random ``disallowed'' place into a network with a percolating rigid
cluster.}
\label{fig:str_perc}
\end{figure}

Similarly to the case of allowed bond insertion, we first look at the
probability that the stressed region percolates. Note that the stressed region
emerging upon bond insertion is always restricted to the rigid cluster
containing the new bond. For this reason, the stressed region can only
percolate if the original network is percolating (i.e., contains a percolating
rigid cluster), and we only need to look at percolating networks. In
Fig.~\ref{fig:str_perc}, we show the probability that in a network in which
rigidity percolates, a percolating stressed region emerges after insertion of
a bond at a randomly chosen disallowed place (not necessarily within the
percolating cluster). We note that apparently, this probability remains finite
(does not go to zero) in the thermodynamic limit.

\begin{figure}
\begin{center}
\includegraphics[width=2.7in]{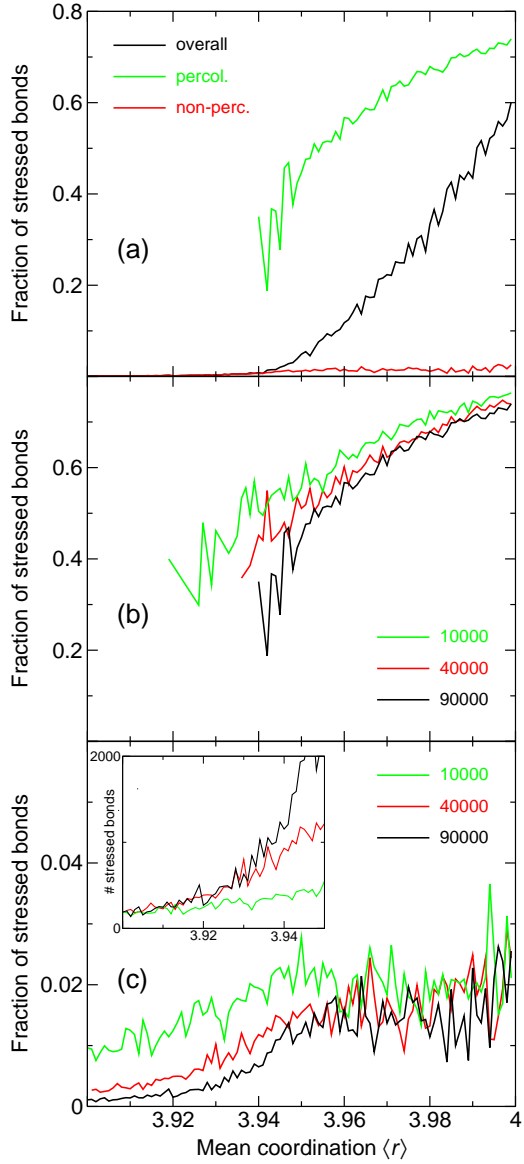}
\end{center}
\caption{The average fraction of stressed bonds in the self-organized network
after a single ``disallowed'' bond is inserted. Panel (a) shows the overall
average, as well as the partial averages for cases when stress does and does
not percolate, for networks of 90000 sites. Panel (b) shows the partial
average for the case when the network percolates, for three different network
sizes. Panel (c) shows the partial average when the network does not percolate,
again, for three different network sizes; in the inset, the average number of
stressed bonds is shown for the same case.}
\label{fig:stress}
\end{figure}

We also look at the sizes of
stressed regions arising upon disallowed bond insertion. The results are in
Fig.~\ref{fig:stress}. Panel (a) shows the overall average fraction of stressed
bonds and the partial
averages for cases when stress does and does not percolate, for networks of
90000 sites; panel (b) gives
the partial average for the case when stress percolates; and panel (c) gives the
partial average for the case when stress does not percolate. We do not make
distinctions based on the {\it rigidity} percolation status of the network.
Again, we see that the size of the stressed region as a finite fraction of the
total number of bonds is only weakly size-independent and thus the affected region
can be macroscopic in the intermediate phase even when there is no percolation.
In the floppy phase, on the other hand, the fraction of stressed bonds decays
fast with increasing network size and as the inset (where the {\it number} of
conversions, rather than the fraction, is plotted) shows, the average number of
conversions is size-independent and the affected region remains finite.

\section{Experimental evidence for the self-organized critical behavior}

\begin{figure}
\begin{center}
\includegraphics[width=2.7in]{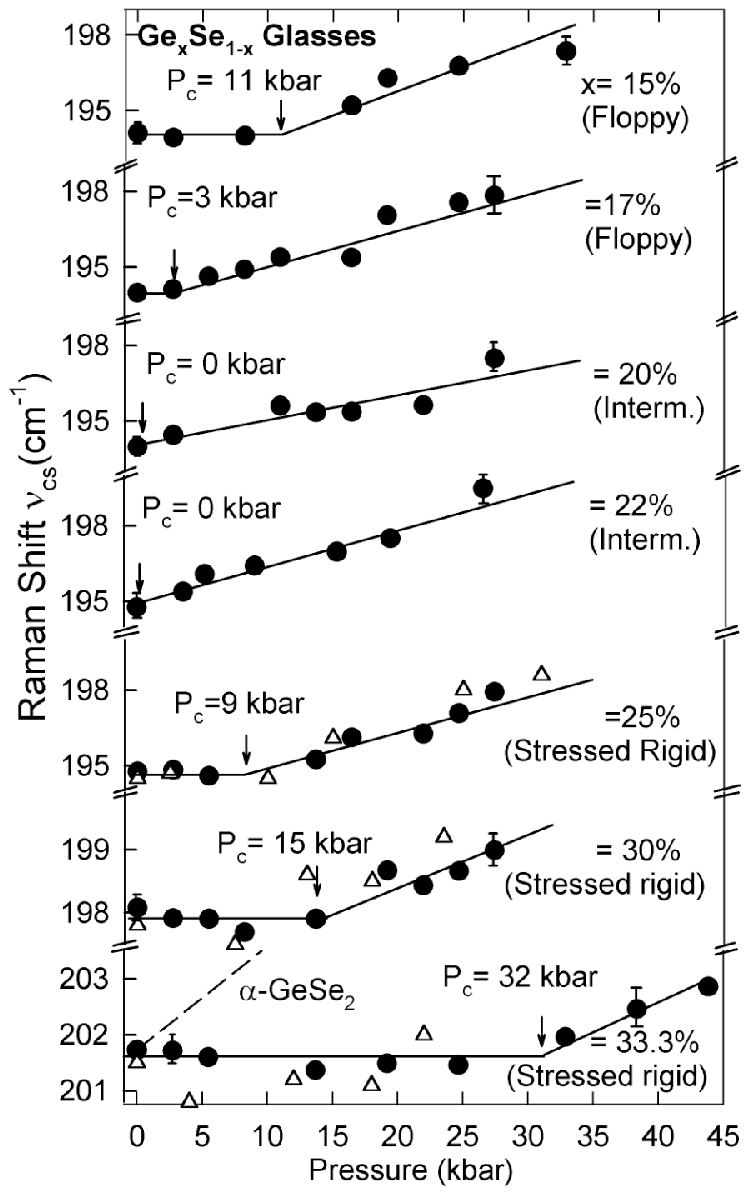}
\end{center}
\caption{Variations in the frequency of CS
tetrahedral units as a function of pressure for
different Ge$_x$Se$_{1-x}$ glasses.  Triangles are
results taken from the work of Murase and Fukunaga
\cite{murase} and filled circles are results of the
work of Wang \emph{et al.}\cite{bool05press}.
The figure is taken from Ref.\cite{bool05press}}
\label{fig:boolchand}
\end{figure}

An atypical response to an external perturbation seen in the intermediate phase has
been observed experimentally~\cite{bool05press}. Raman pressure experiments show that
applying external pressure to a network blueshifts the frequency of
corner-sharing (CS) tetrahedral units only once a certain pressure threshold ($P_c$) value
is reached. $P_c$ is found to be zero within the intermediate phase and non-zero outside the
intermediate phase. These experimental results,
published in Ref.~\cite{bool05press}, are reproduced in Fig.~\ref{fig:boolchand}.

Outside the intermediate phase, the presence of a non-zero threshold for the shift in the
frequency suggests that inhomogeneities cancel the effect, leaving only a broadening
of the peak~\cite{bool05press}. 
On the other hand, we have seen that in the intermediate phase,
a finite fraction of the network is either in a
percolating cluster or in a region that becomes percolating after a few bond
insertions. This region is also stress-free. This can only be achieved by
a precise balancing between constraints and degrees of freedom not just on
average, but on all length scales, in other words, by the network being
{\it homogeneous} because of self-organized criticality. We note that the way rigidity
and stress propagate through a macroscopic fraction of the network in our
simulations is analogous to how stress propagates uniformly when external
pressure is applied in experiments. While the
relation between $P_c$ and inhomogeneity still needs to be fully established,
self-organized criticality in the intermediate phase is consistent with the
observations.

\section{Conclusion}

We have studied the structural and response properties of the intermediate
phase in the phase diagram of rigidity percolation using a model of
self-organization on a 2D triangular network.

We had shown previously that the probability of rigidity percolation in the
intermediate phase increases linearly from zero to one as a function of the
mean coordination $\ra$. At any $\ra$,
there are both percolating and non-percolating networks in the
ensemble. In this paper, we have looked at the properties of both percolating
and non-percolating clusters, the latter separately for percolating and
non-percolating networks. It turns out that at the point at which the
percolating networks first emerge (the rigidity transition), the percolating
cluster takes up about 40\% of the network, unlike the case of the usual
second-order phase transition, where the emerging cluster is fractal and thus
the fraction of bonds belonging to it is zero in the thermodynamic limit.
The size of the percolating cluster and some other characteristics are not
self-averaging, but the distributions are rather narrow, if percolating and
non-percolating networks are considered separately. The
distribution of sizes of non-percolating clusters is exponential at large
sizes in the floppy phase, but power-law for arbitrarily big cluster sizes in
the whole intermediate phase and
not just at the transition. The power-law exponents are different for
non-percolating and percolating networks.

We have also looked at the changes in the rigidity of the network due to a
microscopic perturbation in the form of insertion or removal of a single bond. It
turns out that one bond is often enough to convert a non-percolating network into a
percolating one and vice versa; moreover, when a non-percolating network is turned
into a percolating one, the average size of the resulting percolating cluster is the
same as that for the initially percolating networks. In a sense, the percolating
cluster is ``hidden'' in the non-percolating network and is revealed upon addition of
a single bond. It appears, further, that {\it all} non-percolating networks can be
converted into percolating ones with a finite number of bond additions, with the same
size of the resulting cluster as in networks that percolate without bond addition.
This implies that in the thermodynamic limit, there is no difference between
percolating and non-percolating networks. The intermediate phase then can be thought
of as being the region of the phase diagram where all networks possess a percolating
region that is {\it nearly isostatic}. That is, this region can technically be floppy
or stressed, but the number of floppy modes or redundant constraints causing stress is
negligible in the thermodynamic limit. The lower boundary of the intermediate phase is
then, strictly speaking, not the rigidity percolation transition, but rather the
``near-isostaticity'' percolation transition.

These unusual properties indicate that the intermediate phase is a self-organised
critical phase, with the system staying at the rigidity percolation threshold for the
whole range of mean coordinations. The existence of such a self-organized nearly
isostatic critical phase explains recent pressure experiments~\cite{bool05press}.
Since rigidity percolation occurs on a finite fraction of the network under any
microscropic perturbation, the strain associated with the external pressure is
immediately transferred to a macroscopic fraction of the sample, leading to a shift in
the Raman spectrum.

We can certainly expect other surprises associated with this intermediate
phase in the rigidity phase diagram. But for that, we will likely need to
create more realistic models of the experimental systems.

\section*{Acknowledgments}

The authors are supported in part by the Fonds qu\'eb\'ecois de la recherche sur la
nature et les technologies (FQRNT), the Natural Sciences and Engineering Research
Council of Canada (NSERC) and the Canada Research Chair program. We thank the
R\'eseau qu\'eb\'ecois de calcul de haute performance (RQCHP) for generous allocation
of computational resources.

\section*{Appendix: Relation between the probability of percolation in the
intermediate phase and the probability of inducing percolation by single
bond insertion}

Numerically, the probability that there exists a place in a non-percolating
network such that insertion of a bond at that place causes percolation is the
same linear function in the intermediate phase as the probability of rigidity
percolation (see Fig.~\ref{fig:ins2}). Here we show that this equality is a
necessary consequence of the latter quantity being finite (between zero and
one) and continuous throughout the intermediate phase.

Given the probability of percolation among networks with $B$ bonds, we can
calculate this probability for networks with $B+1$ bonds. The method is
similar to that used to calculate the bond-configurational entropy of
self-organized networks in our previous paper~\cite{chubynsky06}. Suppose there
are $N_{\rm bc}(B)$ networks (or bond configurations) with $B$ bonds that
are stress-free. If the probability of percolation is $s(B)$, then
$N_p(B)=s(B)N_{\rm bc}(B)$ of these are percolating and the rest,
$N_n(B)=[1-s(B)]N_{\rm bc}(B)$, non-percolating.
Also suppose there are on average $N_{\rm ap}(B)$ allowed places to insert a
bond in percolating networks and $N_{\rm an}(B)$ in non-percolating networks and the
probability that a non-percolating network becomes percolating upon a
{\it random} bond insertion is $\pi(B)$.
Since all percolating networks will remain percolating upon bond insertion,
then an average percolating network with $B$ bonds will produce $N_{\rm ap}(B)$
different percolating networks. On the other hand, only a fraction $\pi(B)$ of
non-percolating networks will become percolating, so an average
non-percolating network will produce $\pi(B)N_{\rm an}(B)$ percolating networks and
$[1-\pi(B)]N_{\rm an}(B)$ non-percolating networks. If we simply multiply these
numbers by the number of networks with $B$ bonds of each kind, this will not
produce the correct count of percolating and non-percolating networks with
$B+1$ bonds, because each such network can be obtained in many different ways.
Specifically, each network with $B+1$ bonds can be produced by bond insertion
from as many different networks with $B$ bonds as there are the latter that can
be obtained by bond {\it removal} from the former. This number is always
$B+1$, since removal of every bond will produce a distinct network and all of
them are stress-free. So the count has to be divided by $B+1$, and we get for
the number of percolating networks with $B+1$ bonds,
\begin{eqnarray}
N_p(B+1)&=&\nonumber\\
& &\hspace{-2cm}\frac{N_p(B)N_{\rm ap}(B)+N_n(B)\pi(B)N_{\rm an}(B)}{B+1},
\end{eqnarray}
and for the number of non-percolating networks with $B+1$ bonds,
\begin{equation}
N_n(B+1)=\frac{N_n(B)[1-\pi(B)]N_{\rm an}(B)}{B+1}.
\end{equation}
Then the probability of percolation for a network with $B+1$ bonds is
\begin{eqnarray}
s(B+1)&=&\frac{N_p(B+1)}{N_p(B+1)+N_n(B+1)}\\
& &\hspace{-2cm}=\frac{N_p(B)N_{\rm ap}(B)+N_n(B)\pi(B)N_{\rm an}(B)}
{N_p(B)N_{\rm ap}(B)+N_n(B)N_{\rm an}(B)}.\nonumber
\end{eqnarray}
Note that since the probability of percolation changes continuously, the
difference between $s(B+1)$ and $s(B)$ is $O(1/N)$, where $N$ is the network
size. Neglecting terms that are $O(1/N)$, we should equate
$s(B+1)=s(B)$ and then, omitting the argument $B$ for brevity and using
$N_p/(N_n+N_p)=s$,
\begin{equation}
s=\frac{s N_{\rm ap} + \pi (1-s) N_{\rm an}}{s N_{\rm ap} + (1-s) N_{\rm an}},
\end{equation}
or
\begin{equation}
s=\pi\frac{N_{\rm an}}{N_{\rm an}-N_{\rm ap}}.\label{eq:s}
\end{equation}
Now, as we have seen, networks that become percolating after bond insertion are
typical, generic percolating self-organized networks. So we can expect that on
average, when a bond is inserted into a non-percolating network and it becomes
percolating, the change in the number of allowed bonds is
$N_{\rm an}-N_{\rm ap}$. But bonds that become disallowed are exactly those that
are within the percolating cluster, and, according to our arguments at the
beginning of Section~\ref{sec:pertur}, these and only these bonds will create
the percolating cluster and cause percolation (indeed, if insertion of any
bond creates a percolating cluster, then insertion of any bond outside the
region that turns into the percolating cluster will not cause percolation
because the cluster that will form cannot overlap with this region, other than over
a single pre-existing cluster, and thus
cannot percolate). So, if for a given network making it percolating after one
bond insertion is at all possible, then the probability that percolation will
occur after {\it random} insertion is equal to the fraction of allowed bonds
within the percolating cluster, or $(N_{\rm an}-N_{\rm ap})/N_{\rm an}$, and
then the right-hand side of Eq.~(\ref{eq:s}) is exactly the probability that
percolation after one bond insertion is possible, and the proof is complete.

There are some caveats in this proof, related to the fact that, as we
have seen, many quantities in the intermediate phase are not self-averaging.
Because of this, for instance, if fractions of allowed bonds are not strictly
deterministic, there can be correlations between the number of allowed bonds in
a network and this number in the percolating network created after bond
insertion. For instance, the latter may tend to be above average when the
former is below average. Then the average change in the number of allowed bonds
may differ somewhat from $N_{\rm an}-N_{\rm ap}$, which is the difference of
averages. In fact, the very statement that networks obtained by bond insertion
can be considered typical needs to be treated with caution. Indeed, if all
quantities associated with self-organized networks were self-averaging, then,
in order to change the values of these quantities by inserting a bond, one
would need to upset the balance so badly as to create a finite chance of seeing
networks with properties that at equilibrium have probability zero. Such bias
after inserting just a single bond cannot happen, except in very special cases.
But the situation changes when there is no self-averaging: if the distribution
of values of a certain quantity has a finite width, then any bias can change
the average. For this reason, it is possible that both the equality proved
above and the statement that any quantities obtained for networks that become
percolating after inserting a bond are the same on average as for originally
percolating networks (as demonstrated in Figs.~\ref{fig:clst_compare} and
\ref{fig:backnforth}) are, in fact, only approximate. In any case, deviations
(if any) are very small.


\begin{thebibliography}{33}
\expandafter\ifx\csname natexlab\endcsname\relax\def\natexlab#1{#1}\fi
\expandafter\ifx\csname bibnamefont\endcsname\relax
  \def\bibnamefont#1{#1}\fi
\expandafter\ifx\csname bibfnamefont\endcsname\relax
  \def\bibfnamefont#1{#1}\fi
\expandafter\ifx\csname citenamefont\endcsname\relax
  \def\citenamefont#1{#1}\fi
\expandafter\ifx\csname url\endcsname\relax
  \def\url#1{\texttt{#1}}\fi
\expandafter\ifx\csname urlprefix\endcsname\relax\def\urlprefix{URL }\fi
\providecommand{\bibinfo}[2]{#2}
\providecommand{\eprint}[2][]{\url{#2}}

\bibitem[{\citenamefont{Thorpe}(1983)}]{thorpe83}
\bibinfo{author}{\bibfnamefont{M.~F.} \bibnamefont{Thorpe}},
  \bibinfo{journal}{J. Non-Cryst. Solids} \textbf{\bibinfo{volume}{57}},
  \bibinfo{pages}{355} (\bibinfo{year}{1983}).

\bibitem[{\citenamefont{Phillips}(1979)}]{phillips79}
\bibinfo{author}{\bibfnamefont{J.~C.} \bibnamefont{Phillips}},
  \bibinfo{journal}{J. Non-Cryst. Solids} \textbf{\bibinfo{volume}{34}},
  \bibinfo{pages}{153} (\bibinfo{year}{1979}).

\bibitem[{\citenamefont{Thorpe et~al.}(2000)\citenamefont{Thorpe, Jacobs,
  Chubynsky, and Phillips}}]{thorpe00}
\bibinfo{author}{\bibfnamefont{M.~F.} \bibnamefont{Thorpe}},
  \bibinfo{author}{\bibfnamefont{D.~J.} \bibnamefont{Jacobs}},
  \bibinfo{author}{\bibfnamefont{M.~V.} \bibnamefont{Chubynsky}},
  \bibnamefont{and} \bibinfo{author}{\bibfnamefont{J.~C.}
  \bibnamefont{Phillips}}, \bibinfo{journal}{J. Non-Cryst. Solids}
  \textbf{\bibinfo{volume}{266-269}}, \bibinfo{pages}{859}
  (\bibinfo{year}{2000}).

\bibitem[{\citenamefont{Boolchand et~al.}(1999)\citenamefont{Boolchand, Feng,
  Selvanathan, and Bresser}}]{travbool}
\bibinfo{author}{\bibfnamefont{P.}~\bibnamefont{Boolchand}},
  \bibinfo{author}{\bibfnamefont{X.}~\bibnamefont{Feng}},
  \bibinfo{author}{\bibfnamefont{D.}~\bibnamefont{Selvanathan}},
  \bibnamefont{and} \bibinfo{author}{\bibfnamefont{W.~J.}
  \bibnamefont{Bresser}}, in \emph{\bibinfo{booktitle}{Rigidity Theory and
  Applications}}, edited by \bibinfo{editor}{\bibfnamefont{M.~F.}
  \bibnamefont{Thorpe}} \bibnamefont{and} \bibinfo{editor}{\bibfnamefont{P.~M.}
  \bibnamefont{Duxbury}} (\bibinfo{publisher}{Kluwer Academic/Plenum
  Publishers, New York}, \bibinfo{year}{1999}), p. \bibinfo{pages}{279}.

\bibitem[{\citenamefont{Thorpe et~al.}(1999)\citenamefont{Thorpe, Jacobs,
  Chubynsky, and Rader}}]{travthorpe}
\bibinfo{author}{\bibfnamefont{M.~F.} \bibnamefont{Thorpe}},
  \bibinfo{author}{\bibfnamefont{D.~J.} \bibnamefont{Jacobs}},
  \bibinfo{author}{\bibfnamefont{N.~V.} \bibnamefont{Chubynsky}},
  \bibnamefont{and} \bibinfo{author}{\bibfnamefont{A.~J.} \bibnamefont{Rader}},
  in \emph{\bibinfo{booktitle}{Rigidity Theory and Applications}}, edited by
  \bibinfo{editor}{\bibfnamefont{M.~F.} \bibnamefont{Thorpe}} \bibnamefont{and}
  \bibinfo{editor}{\bibfnamefont{P.~M.} \bibnamefont{Duxbury}}
  (\bibinfo{publisher}{Kluwer Academic/Plenum Publishers, New York},
  \bibinfo{year}{1999}), p. \bibinfo{pages}{239}.

\bibitem[{\citenamefont{Jacobs et~al.}(2001)\citenamefont{Jacobs, Rader, Kuhn,
  and Thorpe}}]{proteins01}
\bibinfo{author}{\bibfnamefont{D.~J.} \bibnamefont{Jacobs}},
  \bibinfo{author}{\bibfnamefont{A.~J.} \bibnamefont{Rader}},
  \bibinfo{author}{\bibfnamefont{L.~A.} \bibnamefont{Kuhn}}, \bibnamefont{and}
  \bibinfo{author}{\bibfnamefont{M.~F.} \bibnamefont{Thorpe}},
  \bibinfo{journal}{Proteins} \textbf{\bibinfo{volume}{44}},
  \bibinfo{pages}{150} (\bibinfo{year}{2001}).

\bibitem[{\citenamefont{Selvanathan et~al.}(2000)\citenamefont{Selvanathan,
  Bresser, and Boolchand}}]{bool00sise}
\bibinfo{author}{\bibfnamefont{D.}~\bibnamefont{Selvanathan}},
  \bibinfo{author}{\bibfnamefont{W.~J.} \bibnamefont{Bresser}},
  \bibnamefont{and}
  \bibinfo{author}{\bibfnamefont{P.}~\bibnamefont{Boolchand}},
  \bibinfo{journal}{Phys. Rev. B} \textbf{\bibinfo{volume}{61}},
  \bibinfo{pages}{15061} (\bibinfo{year}{2000}).

\bibitem[{\citenamefont{Boolchand
  et~al.}(2001{\natexlab{a}})\citenamefont{Boolchand, Georgiev, and
  Goodman}}]{bool01rev}
\bibinfo{author}{\bibfnamefont{P.}~\bibnamefont{Boolchand}},
  \bibinfo{author}{\bibfnamefont{D.~G.} \bibnamefont{Georgiev}},
  \bibnamefont{and} \bibinfo{author}{\bibfnamefont{B.}~\bibnamefont{Goodman}},
  \bibinfo{journal}{J. Optoelectron. Adv. Mater.} \textbf{\bibinfo{volume}{3}},
  \bibinfo{pages}{703} (\bibinfo{year}{2001}{\natexlab{a}}).

\bibitem[{\citenamefont{Boolchand
  et~al.}(2001{\natexlab{b}})\citenamefont{Boolchand, Feng, and
  Bresser}}]{bool01gese}
\bibinfo{author}{\bibfnamefont{P.}~\bibnamefont{Boolchand}},
  \bibinfo{author}{\bibfnamefont{X.}~\bibnamefont{Feng}}, \bibnamefont{and}
  \bibinfo{author}{\bibfnamefont{W.~J.} \bibnamefont{Bresser}},
  \bibinfo{journal}{J. Non-Cryst. Solids} \textbf{\bibinfo{volume}{293}},
  \bibinfo{pages}{348} (\bibinfo{year}{2001}{\natexlab{b}}).

\bibitem[{\citenamefont{Boolchand
  et~al.}(2002{\natexlab{a}})\citenamefont{Boolchand, Georgiev, and
  Micoulaut}}]{bool02rev}
\bibinfo{author}{\bibfnamefont{P.}~\bibnamefont{Boolchand}},
  \bibinfo{author}{\bibfnamefont{D.~G.} \bibnamefont{Georgiev}},
  \bibnamefont{and}
  \bibinfo{author}{\bibfnamefont{M.}~\bibnamefont{Micoulaut}},
  \bibinfo{journal}{J. Optoelectron. Adv. Mater.} \textbf{\bibinfo{volume}{4}},
  \bibinfo{pages}{823} (\bibinfo{year}{2002}{\natexlab{a}}).

\bibitem[{\citenamefont{Boolchand
  et~al.}(2002{\natexlab{b}})\citenamefont{Boolchand, Georgiev, Qu, Wang, Cai,
  and Chakravarty}}]{bool02cr}
\bibinfo{author}{\bibfnamefont{P.}~\bibnamefont{Boolchand}},
  \bibinfo{author}{\bibfnamefont{D.~G.} \bibnamefont{Georgiev}},
  \bibinfo{author}{\bibfnamefont{T.}~\bibnamefont{Qu}},
  \bibinfo{author}{\bibfnamefont{F.}~\bibnamefont{Wang}},
  \bibinfo{author}{\bibfnamefont{L.}~\bibnamefont{Cai}}, \bibnamefont{and}
  \bibinfo{author}{\bibfnamefont{S.}~\bibnamefont{Chakravarty}},
  \bibinfo{journal}{C.R. Chimie} \textbf{\bibinfo{volume}{5}},
  \bibinfo{pages}{713} (\bibinfo{year}{2002}{\natexlab{b}}).

\bibitem[{\citenamefont{Georgiev et~al.}(2003)\citenamefont{Georgiev,
  Boolchand, Eckert, Micoulaut, and Jackson}}]{bool03pse}
\bibinfo{author}{\bibfnamefont{D.~G.} \bibnamefont{Georgiev}},
  \bibinfo{author}{\bibfnamefont{P.}~\bibnamefont{Boolchand}},
  \bibinfo{author}{\bibfnamefont{H.}~\bibnamefont{Eckert}},
  \bibinfo{author}{\bibfnamefont{M.}~\bibnamefont{Micoulaut}},
  \bibnamefont{and} \bibinfo{author}{\bibfnamefont{K.}~\bibnamefont{Jackson}},
  \bibinfo{journal}{Europhys. Lett.} \textbf{\bibinfo{volume}{62}},
  \bibinfo{pages}{49} (\bibinfo{year}{2003}).

\bibitem[{\citenamefont{Qu et~al.}(2003)\citenamefont{Qu, Georgiev, Boolchand,
  and Micoulaut}}]{bool03geasse}
\bibinfo{author}{\bibfnamefont{T.}~\bibnamefont{Qu}},
  \bibinfo{author}{\bibfnamefont{D.~G.} \bibnamefont{Georgiev}},
  \bibinfo{author}{\bibfnamefont{P.}~\bibnamefont{Boolchand}},
  \bibnamefont{and}
  \bibinfo{author}{\bibfnamefont{M.}~\bibnamefont{Micoulaut}}, in
  \emph{\bibinfo{booktitle}{Mat. Res. Soc. Symp. Proc. Vol. 754}}, edited by
  \bibinfo{editor}{\bibfnamefont{T.}~\bibnamefont{Egami}},
  \bibinfo{editor}{\bibfnamefont{A.~L.} \bibnamefont{Greer}},
  \bibinfo{editor}{\bibfnamefont{A.}~\bibnamefont{Inoue}}, \bibnamefont{and}
  \bibinfo{editor}{\bibfnamefont{S.}~\bibnamefont{Ranganathan}}
  (\bibinfo{year}{2003}), p. \bibinfo{pages}{CC8.1}.

\bibitem[{\citenamefont{Vempati and Boolchand}(2004)}]{bool04geps}
\bibinfo{author}{\bibfnamefont{U.}~\bibnamefont{Vempati}} \bibnamefont{and}
  \bibinfo{author}{\bibfnamefont{P.}~\bibnamefont{Boolchand}},
  \bibinfo{journal}{J. Phys.: Condens. Mat.} \textbf{\bibinfo{volume}{16}},
  \bibinfo{pages}{S5121} (\bibinfo{year}{2004}).

\bibitem[{\citenamefont{Chakravarty et~al.}(2005)\citenamefont{Chakravarty,
  Georgiev, Boolchand, and Micoulaut}}]{bool05pgese}
\bibinfo{author}{\bibfnamefont{S.}~\bibnamefont{Chakravarty}},
  \bibinfo{author}{\bibfnamefont{D.~G.} \bibnamefont{Georgiev}},
  \bibinfo{author}{\bibfnamefont{P.}~\bibnamefont{Boolchand}},
  \bibnamefont{and}
  \bibinfo{author}{\bibfnamefont{M.}~\bibnamefont{Micoulaut}},
  \bibinfo{journal}{J. Phys.: Condens. Mat.} \textbf{\bibinfo{volume}{17}},
  \bibinfo{pages}{L1} (\bibinfo{year}{2005}).

\bibitem[{\citenamefont{\v{C}erno\v{s}kov\'a
  et~al.}(2005)\citenamefont{\v{C}erno\v{s}kov\'a, Qu, Mamedov,
  \v{C}erno\v{s}ek, Holubov\'a, and Boolchand}}]{bool05geass}
\bibinfo{author}{\bibfnamefont{E.}~\bibnamefont{\v{C}erno\v{s}kov\'a}},
  \bibinfo{author}{\bibfnamefont{T.}~\bibnamefont{Qu}},
  \bibinfo{author}{\bibfnamefont{S.}~\bibnamefont{Mamedov}},
  \bibinfo{author}{\bibnamefont{Z.}~\bibnamefont{\v{C}erno\v{s}ek}},
  \bibinfo{author}{\bibfnamefont{J.}~\bibnamefont{Holubov\'a}},
  \bibnamefont{and}
  \bibinfo{author}{\bibfnamefont{P.}~\bibnamefont{Boolchand}},
  \bibinfo{journal}{J. Phys. Chem. Solids} \textbf{\bibinfo{volume}{66}},
  \bibinfo{pages}{185} (\bibinfo{year}{2005}).

\bibitem[{\citenamefont{Qu and Boolchand}(2005)}]{bool05geass2}
\bibinfo{author}{\bibfnamefont{T.}~\bibnamefont{Qu}} \bibnamefont{and}
  \bibinfo{author}{\bibfnamefont{P.}~\bibnamefont{Boolchand}},
  \bibinfo{journal}{Phil. Mag.} \textbf{\bibinfo{volume}{66}},
  \bibinfo{pages}{875} (\bibinfo{year}{2005}).

\bibitem[{\citenamefont{Wang et~al.}(2005)\citenamefont{Wang, Mamedov,
  Boolchand, Goodman, and Chandrasekhar}}]{bool05press}
\bibinfo{author}{\bibfnamefont{F.}~\bibnamefont{Wang}},
  \bibinfo{author}{\bibfnamefont{S.}~\bibnamefont{Mamedov}},
  \bibinfo{author}{\bibfnamefont{P.}~\bibnamefont{Boolchand}},
  \bibinfo{author}{\bibfnamefont{B.}~\bibnamefont{Goodman}}, \bibnamefont{and}
  \bibinfo{author}{\bibfnamefont{M.}~\bibnamefont{Chandrasekhar}},
  \bibinfo{journal}{Phys. Rev. B} \textbf{\bibinfo{volume}{71}},
  \bibinfo{pages}{174201} (\bibinfo{year}{2005}).

\bibitem[{\citenamefont{Vaills et~al.}(2005)\citenamefont{Vaills, Qu,
  Micoulaut, Chaimbault, and Boolchand}}]{bool05sil}
\bibinfo{author}{\bibfnamefont{Y.}~\bibnamefont{Vaills}},
  \bibinfo{author}{\bibfnamefont{T.}~\bibnamefont{Qu}},
  \bibinfo{author}{\bibfnamefont{M.}~\bibnamefont{Micoulaut}},
  \bibinfo{author}{\bibfnamefont{F.}~\bibnamefont{Chaimbault}},
  \bibnamefont{and}
  \bibinfo{author}{\bibfnamefont{P.}~\bibnamefont{Boolchand}},
  \bibinfo{journal}{J. Phys.: Condens. Mat.} \textbf{\bibinfo{volume}{17}},
  \bibinfo{pages}{4889} (\bibinfo{year}{2005}).

\bibitem[{\citenamefont{Lucovsky and Phillips}(2004)}]{lucov4}
\bibinfo{author}{\bibfnamefont{G.}~\bibnamefont{Lucovsky}} \bibnamefont{and}
  \bibinfo{author}{\bibfnamefont{J.~C.} \bibnamefont{Phillips}},
  \bibinfo{journal}{Appl. Phys. A} \textbf{\bibinfo{volume}{78}},
  \bibinfo{pages}{453} (\bibinfo{year}{2004}).

\bibitem[{\citenamefont{Thorpe and Chubynsky}(2001)}]{czech01}
\bibinfo{author}{\bibfnamefont{M.~F.} \bibnamefont{Thorpe}} \bibnamefont{and}
  \bibinfo{author}{\bibfnamefont{M.~V.} \bibnamefont{Chubynsky}}, in
  \emph{\bibinfo{booktitle}{Properties and Applications of Amorphous
  Materials}}, edited by \bibinfo{editor}{\bibfnamefont{M.~F.}
  \bibnamefont{Thorpe}} \bibnamefont{and}
  \bibinfo{editor}{\bibfnamefont{L.}~\bibnamefont{Tich\'y}}
  (\bibinfo{publisher}{Kluwer Academic, Dordrecht}, \bibinfo{year}{2001}), NATO
  Science Series, II. Mathematics, Physics and Chemistry, vol. 9,
  p.~\bibinfo{pages}{61}.

\bibitem[{\citenamefont{Micoulaut}(2002)}]{micoulaut02}
\bibinfo{author}{\bibfnamefont{M.}~\bibnamefont{Micoulaut}},
  \bibinfo{journal}{Europhys. Lett.} \textbf{\bibinfo{volume}{58}},
  \bibinfo{pages}{830} (\bibinfo{year}{2002}).

\bibitem[{\citenamefont{Micoulaut and Phillips}(2003)}]{micoulaut03}
\bibinfo{author}{\bibfnamefont{M.}~\bibnamefont{Micoulaut}} \bibnamefont{and}
  \bibinfo{author}{\bibfnamefont{J.~C.} \bibnamefont{Phillips}},
  \bibinfo{journal}{Phys. Rev. B} \textbf{\bibinfo{volume}{67}},
  \bibinfo{pages}{104204} (\bibinfo{year}{2003}).

\bibitem[{\citenamefont{Barr\'e et~al.}(2005)\citenamefont{Barr\'e, Bishop,
  Lookman, and Saxena}}]{barre}
\bibinfo{author}{\bibfnamefont{J.}~\bibnamefont{Barr\'e}},
  \bibinfo{author}{\bibfnamefont{A.~R.} \bibnamefont{Bishop}},
  \bibinfo{author}{\bibfnamefont{T.}~\bibnamefont{Lookman}}, \bibnamefont{and}
  \bibinfo{author}{\bibfnamefont{A.}~\bibnamefont{Saxena}},
  \bibinfo{journal}{Phys. Rev. Lett.} \textbf{\bibinfo{volume}{94}},
  \bibinfo{pages}{208701} (\bibinfo{year}{2005}).

\bibitem[{\citenamefont{Bak et~al.}(1988)\citenamefont{Bak, Tang, and
  Wiesenfeld}}]{bak2}
\bibinfo{author}{\bibfnamefont{P.}~\bibnamefont{Bak}},
  \bibinfo{author}{\bibfnamefont{C.}~\bibnamefont{Tang}}, \bibnamefont{and}
  \bibinfo{author}{\bibfnamefont{K.}~\bibnamefont{Wiesenfeld}},
  \bibinfo{journal}{Phys. Rev. A} \textbf{\bibinfo{volume}{38}},
  \bibinfo{pages}{364} (\bibinfo{year}{1988}).

\bibitem[{\citenamefont{Maxwell}(1864)}]{maxwell}
\bibinfo{author}{\bibfnamefont{J.~C.} \bibnamefont{Maxwell}},
  \bibinfo{journal}{Philos. Mag.} \textbf{\bibinfo{volume}{27}},
  \bibinfo{pages}{294} (\bibinfo{year}{1864}).

\bibitem[{\citenamefont{Jacobs and Thorpe}(1995)}]{jacobs95}
\bibinfo{author}{\bibfnamefont{D.~J.} \bibnamefont{Jacobs}} \bibnamefont{and}
  \bibinfo{author}{\bibfnamefont{M.~F.} \bibnamefont{Thorpe}},
  \bibinfo{journal}{Phys. Rev. Lett.} \textbf{\bibinfo{volume}{75}},
  \bibinfo{pages}{4051} (\bibinfo{year}{1995}).

\bibitem[{\citenamefont{Jacobs and Hendrickson}(1997)}]{jacobs97}
\bibinfo{author}{\bibfnamefont{D.~J.} \bibnamefont{Jacobs}} \bibnamefont{and}
  \bibinfo{author}{\bibfnamefont{B.}~\bibnamefont{Hendrickson}},
  \bibinfo{journal}{J. Comput. Phys.} \textbf{\bibinfo{volume}{137}},
  \bibinfo{pages}{346} (\bibinfo{year}{1997}).

\bibitem[{\citenamefont{Laman}(1970)}]{laman}
\bibinfo{author}{\bibfnamefont{G.}~\bibnamefont{Laman}}, \bibinfo{journal}{J.
  Engrg. Math.} \textbf{\bibinfo{volume}{4}}, \bibinfo{pages}{331}
  (\bibinfo{year}{1970}).

\bibitem[{\citenamefont{Chubynsky et~al.}(2006)\citenamefont{Chubynsky,
  Bri\`ere, and Mousseau}}]{chubynsky06}
\bibinfo{author}{\bibfnamefont{M.~V.} \bibnamefont{Chubynsky}},
  \bibinfo{author}{\bibfnamefont{M.-A.} \bibnamefont{Bri\`ere}},
  \bibnamefont{and} \bibinfo{author}{\bibfnamefont{N.}~\bibnamefont{Mousseau}},
  \bibinfo{journal}{Phys. Rev. E} \textbf{\bibinfo{volume}{74}},
  \bibinfo{pages}{016116} (\bibinfo{year}{2006}).

\bibitem[{\citenamefont{Moukarzel et~al.}(1997)\citenamefont{Moukarzel,
  Duxbury, and Leath}}]{dux97}
\bibinfo{author}{\bibfnamefont{C.}~\bibnamefont{Moukarzel}},
  \bibinfo{author}{\bibfnamefont{P.~M.} \bibnamefont{Duxbury}},
  \bibnamefont{and} \bibinfo{author}{\bibfnamefont{P.~L.} \bibnamefont{Leath}},
  \bibinfo{journal}{Phys. Rev. E} \textbf{\bibinfo{volume}{55}},
  \bibinfo{pages}{5800} (\bibinfo{year}{1997}).

\bibitem[{\citenamefont{Duxbury et~al.}(1999)\citenamefont{Duxbury, Jacobs,
  Thorpe, and Moukarzel}}]{dux99}
\bibinfo{author}{\bibfnamefont{P.~M.} \bibnamefont{Duxbury}},
  \bibinfo{author}{\bibfnamefont{D.~J.} \bibnamefont{Jacobs}},
  \bibinfo{author}{\bibfnamefont{M.~F.} \bibnamefont{Thorpe}},
  \bibnamefont{and}
  \bibinfo{author}{\bibfnamefont{C.}~\bibnamefont{Moukarzel}},
  \bibinfo{journal}{Phys. Rev. E} \textbf{\bibinfo{volume}{59}},
  \bibinfo{pages}{2084} (\bibinfo{year}{1999}).

\bibitem[{\citenamefont{Murase and Fukunaga}(1984)}]{murase}
\bibinfo{author}{\bibfnamefont{K.}~\bibnamefont{Murase}} \bibnamefont{and}
  \bibinfo{author}{\bibfnamefont{T.}~\bibnamefont{Fukunaga}}, in
  \emph{\bibinfo{booktitle}{Optical Effects in Amorphous Semiconductors}},
  edited by \bibinfo{editor}{\bibfnamefont{P.~C.} \bibnamefont{Taylor}}
  \bibnamefont{and} \bibinfo{editor}{\bibfnamefont{S.~G.} \bibnamefont{Bishop}}
  (\bibinfo{publisher}{American Institute of Physics}, \bibinfo{year}{1984}),
  p. \bibinfo{pages}{449}.

\end{thebibliography}
\end{document}